\documentclass[aip,jcp,amsmath,amssymb,amsfonts,reprint,groupedaddress,floatfix]{revtex4-2}
\usepackage{graphicx}
\usepackage{dcolumn}
\usepackage{bm}
\usepackage{placeins}
\usepackage{rotating}
\usepackage{multirow}
\usepackage[dvipsnames,usenames]{xcolor}
\usepackage[english]{babel}
\usepackage{amssymb}
\usepackage{amsmath}
\usepackage[utf8]{inputenc}
\usepackage[normalem]{ulem}
\usepackage{float}
\usepackage{array}
\usepackage{bigints}
\usepackage[colorinlistoftodos]{todonotes}
\usepackage{soul}

\newcommand{\be}{\begin{equation}}
\newcommand{\ee}{\end{equation}}
\newcommand{\bea}{\begin{eqnarray}}
\newcommand{\eea}{\end{eqnarray}}

\newcolumntype{M}[1]{>{\centering\arraybackslash}m{#1}}
\newcolumntype{N}{@{}m{0pt}@{}}

\def\a{\alpha}

\def\g{\gamma}

\def\d{\delta}
\def\D{\Delta}

\def\l{\lambda}
\def\L{\Lambda}

\begin{document}
\widetext

\title{Assessing many-body methods on the potential energy surface of the (H$_2$)$_2$ hydrogen dimer}
\author{Damian Contant}
\homepage{Corresponding author (damian.contant@sorbonne-universite.fr)}
\affiliation{Sorbonne Université, MNHN, UMR CNRS 7590, IMPMC, 4 place Jussieu, 75005 Paris, France}
\author{Michele Casula}
\affiliation{Sorbonne Université, MNHN, UMR CNRS 7590, IMPMC, 4 place Jussieu, 75005 Paris, France}
\author{Maria Hellgren}
\affiliation{Sorbonne Université, MNHN, UMR CNRS 7590, IMPMC, 4 place Jussieu, 75005 Paris, France}

\date{\today}

\begin{abstract}
The anisotropic potential energy surface of the (H$_2$)$_2$ dimer represents a challenging problem for many-body methods. Here, we determine the potential energy curves of five different dimer configurations (T, Z, X, H, L) using the lattice regularized diffusion Monte Carlo (LRDMC) method and a number of approximate functionals within density functional theory (DFT), including advanced orbital-dependent functionals based on the random phase approximation (RPA). We assess their performance in describing the potential wells, bond distances and relative energies. The repulsive potential wall is studied by looking at the relative stability of the different dimer configurations as a function of an applied force acting along the intermolecular axis. It is shown that most functionals within DFT break down at finite compression, even those that give an accurate description around the potential well minima. Only by including exchange within RPA a qualitatively correct description along the entire potential energy curve is obtained. Finally, we discuss these results in the context of solid molecular hydrogen at finite pressures. 
\end{abstract}

\maketitle

\section{\label{sec:Part1} Introduction}
The hydrogen molecule is among the simplest chemical systems. It is often considered in numerical studies for the purpose of testing approximate many-electron approaches. Close to equilibrium, the chemical bonding is strong and effects of correlation play a minor role. However, by increasing the nuclear separation, electrons tend to localize on the ions and an accurate description of static correlation becomes essential. The stretched H$_2$ molecule is, therefore, a simple, yet severe, test case for methods that aim to capture strong correlation.\cite{Fuchs2005,helbig2009,Stella2011,benavides2017towards}

The weak anisotropic interaction between two hydrogen molecules also poses a challenge to approximate methods due to a subtle interplay between van der Waals (vdW), exchange, and electrostatic forces. According to CCSD(T) calculations, the lowest energy (H$_2$)$_2$ configuration is T-shaped with a dissociation energy of 52 K, while the highest energy configuration is linear (L-shaped) with a dissociation energy of 14 K.\cite{Patkowski2008,Lu2013} 
In-between the T and L orientations are additional stationary points: the parallel H-shaped dimer, the crossed X-shaped dimer and the tilted Z-shaped dimer (see Fig.~\ref{fig:Figure1_Main}). The potential barriers between these configurations are, however, small enough for the dimer to always exist in a mixture of different configurations.\cite{Watanabe1964,McKellar1991,Hinde2008}
An isotropic potential energy surface (PES) has been determined experimentally, giving a potential well minimum of 30-40 K, and an estimated binding energy of 4-6 K.\cite{Khan2020}

Besides the interest in studying $\rm H_{2}$ and $\rm (H_{2})_{2}$ as a playground where different types of electronic interactions are present, these systems can be seen as elementary units of a more general interaction, involved in the formation and stability of condensed phases of hydrogen.
At low temperature and ambient pressure, hydrogen molecules form a solid which, similarly to the (H$_2$)$_2$ dimers, consists of essentially freely rotating molecules.\cite{Silvera1980,Mao1994,McMahon2012} At finite pressures, the rotational motion is suppressed and well-defined molecular crystals are observed.\cite{Lorenzana1990,Hanfland1993,Goncharov1998} 
Various high-pressure phase transitions occur, notably the II-III phase transition, in which the hydrogen molecules become subject to strong in-plane interactions, forming a layered structure with polarized H$_2$ units.\cite{Hemley1988,Lorenzana1989,Pickard2007,Akahama2010,Hellgren2022} The H$_2$ molecular bond is, eventually, expected to dissociate with the formation of a metallic atomic phase.\cite{Wigner1935,Ashcroft1968,Morales2015,Monacelli2023} In contrast to the stretched H$_2$ molecule and the weakly interacting (H$_2$)$_2$ dimer, the properties of the crystal phases are determined at high compression, where the nature of chemical bonding and the role of electronic correlation are less explored.\cite{Labet2012,riffet2017} In this context, Diffusion Monte Carlo (DMC) is often used as a benchmark method for its capability of including electron correlation in a very flexible way by preserving a high level of accuracy across diverse interaction regimes.\cite{wagner2016discovering} 
In Refs.~\onlinecite{Hellgren2022,Morales2014}, a number of functionals within density functional theory (DFT) were benchmarked on the high-pressure phases of solid hydrogen. The results obtained proved to be very sensitive to the choice of functional; for example, the predicted transition pressures of the II-III phase transition are spread within an interval as large as 200 GPa, and it is not always clear why some approximations work better than others. 

In order to gain further insights, we have, in this work, carried out an extensive analysis of the performance of different methods on the PES of the isolated (H$_2$)$_2$ dimer: at equilibrium and at finite compression. Accurate reference results have been generated with the DMC method in the lattice regularized flavor (LRDMC).\cite{Casula2005} Several methods/approximations have been tested, specifically a number of relevant functionals within DFT (BLYP,\cite{Becke1988,Lee1988} B3LYP,\cite{becke1992density,becke1993} B3LYP-D3,\cite{grimme2010consistent} PBE,\cite{Burke1996,PhysRevLett.78.1396} PBE0,\cite{Adamo1999} vdW-DF,\cite{dion2004van,Thonhauser2015} vdW-DF2\cite{lee2010higher}), the Hartree-Fock (HF) method,\cite{HF_paper} and the orbital-dependent RPA (random phase approximation) and RPA + exchange (RPAx) functionals.\cite{Kummel2008,ren2012random,eshuis2012electron,Hellgren2010,hesselmann2010,Bleiziffer2015,Hellgren2016,Hellgren2018,Hellgren2021} 

The paper is organized as follows. In Sec.~\ref{sec:Part2}, we introduce and discuss the employed theoretical methods, in particular the more advanced DMC and RPA-based methods. We also provide the computational details of our calculations. In Sec.~\ref{subsec:Part3A}, we discuss the results (dissociation energies, bond distances and relative energies) obtained on five different (H$_2$)$_2$ dimer configurations (T, Z, X, H, L) around equilibrium geometry. In Sec.~\ref{subsec:Part3B}, we study the evolution of the results under the application of a compressive force acting along the intermolecular axis. Finally, we present our conclusions in Sec.~\ref{sec:Part4}. 

\section{\label{sec:Part2} Theory and computational details}
In this Section, we briefly review the many-body methods employed in this work, focusing on the DMC method and RPA-based approximations. We also give a comprehensive account of the computational details for the full set of calculations presented in Sec.~\ref{sec:Part3}. 
\subsection{\label{subsec:Part2A} Quantum Monte Carlo}
Continuum quantum Monte Carlo (QMC) methods employed in this work rely on determining the best possible representation of the unknown ground state wave function of the system. At the variational Monte Carlo (VMC) level, the wave function $\Psi_{\rm VMC}$ is expressed as a product of a Jastrow factor and a determinantal part, as follows:
\begin{equation}
\displaystyle \Psi_{\rm VMC}(\mathbf{R}) = e^{-J(\mathbf{R})} \det[ \Phi (\mathbf{r}_i^\uparrow, \mathbf{r}_j^\downarrow) ],
\label{eq:QMC_wave function}
\end{equation}
where $J$ is the Jastrow function, $\Phi$ is a geminal function representing the spatial part of an electron singlet,\cite{Casula2003,bajdich2006pfaffian} $\mathbf{R}$ is an all-electron coordinate, and $\mathbf{r}_i^\sigma$ is the position of the $i$-th electron with spin $\sigma$ included in $\mathbf{R}$.

The geminal $\Phi$ is expanded on a contracted Gaussian basis set made of 6 geminal embedded orbitals (GEO) of hybrid character,\cite{sorella2015geminal} developed on a $4s 2p 1d$ primitive basis set. The geminal expansion covers a variational freedom ranging from the single Slater determinant (SD), yielded by its lowest-ranked expression, to a multi-determinant expansion with zero seniority, obtained within the antisymmetrized geminal power (AGP) ansatz.\cite{Coleman1965,bytautas2011seniority} In this work, we exploited the single Slater representation of the geminal and investigated the geminal's capability of describing static correlation through an expansion beyond the single Slater determinant.\cite{casula2004correlated,zen2014static} In the constrained SD form, $\Phi$ has been built starting from Kohn-Sham orbitals generated by DFT within the local density approximation (LDA).\cite{Kohn1965} In a subsequent optimization step, these one-body orbitals have been fully relaxed within the Jastrow correlated SD ansatz.

The Jastrow function $J$ has been developed by including both the exponential one-body function $u_{1b}$ ($u_{1b}(r_{iN}) = (1-\exp(-a r_{iN}))/a$, where $r_{iN}$ is the electron-nucleus distance and $a$ is a variational parameter) as well as the Pad\'e two-body function $u_{2b}$ ($u_{2b}(r_{ij}) = 0.5 r_{ij}/(1+b r_{ij})$, where $r_{ij}$ is the electron-electron distance and $b$ is a variational parameter). These two functions respectively fulfill the electron-nucleus cusp condition and the unlike-spin electron-electron cusp condition. For describing higher order correlations, a further term is added to $J$ including up to 4-body (i.e. 2 electrons and 2 nuclei) correlations. It is developed in a way similar to the geminal function $\Phi$, this time expanded on a $2s 2p 1d$ basis set made of Gaussian orbitals for all elements except for the second $s$ component and for the radial part of the second $p$ component of the basis set, for which Gaussian times $r^2$ orbitals have instead been chosen. A more comprehensive description of the Jastrow correlated wave function presented in Eq.~\ref{eq:QMC_wave function} can be found in Ref.~\onlinecite{Nakano2020}.
\begin{figure}[t]
\begin{center}
\includegraphics[scale=0.55,angle=0]{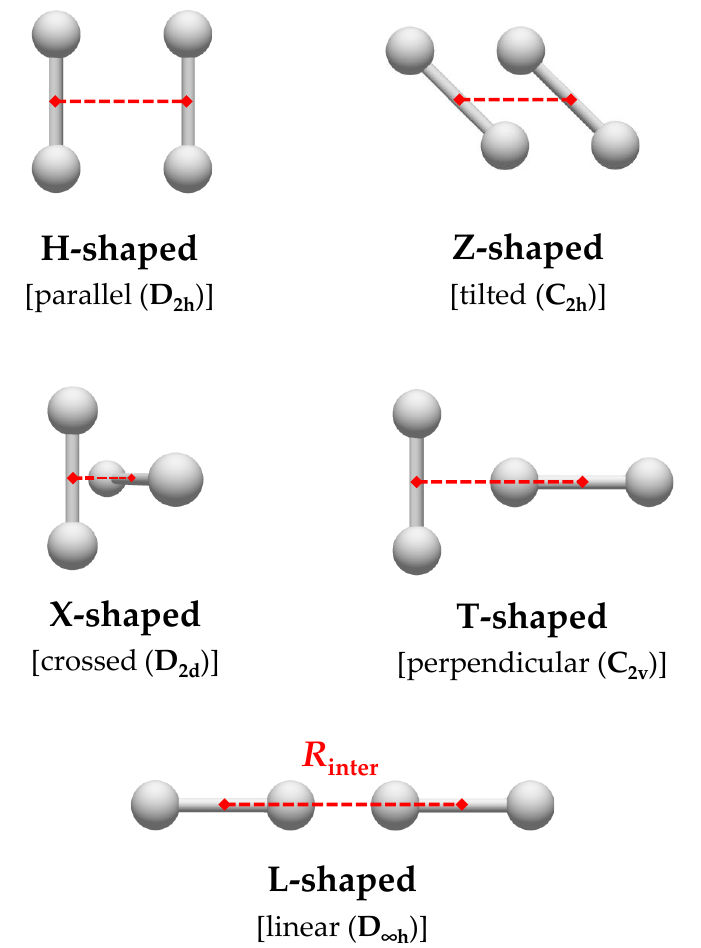}
\caption{\label{fig:Figure1_Main} Schematic representation of the five different (H$_2$)$_2$ configurations considered in this study. The intermolecular distance $R_{\rm inter}$ is defined as the distance between the center of mass of the H$_2$ molecules.}
\end{center}
\end{figure}
Once the LDA orbitals are loaded into the correlated wave function, $\Psi_{\rm VMC}$ is then optimized by energy minimization within the stochastic reconfiguration (SR) scheme.\cite{Sorella_opt1, Becca2017} The Jastrow function $J$ is first relaxed in terms of both the linear coefficients and the basis set exponents. For the latter, the SR approach is supplemented by the information on the energy curvature through the linear method.\cite{Umrigar_opt} We thus obtain the best Jastrow correlated wave function with frozen LDA orbitals (JDFT). We can then proceed further by relaxing both the Jastrow and the determinantal part, relying again on the linear method as energy minimizer. After this step, we get the optimal one-body orbitals in presence of Jastrow correlations. We thus lose any dependence on the LDA starting point and get the best $\Psi_{\rm VMC}$ in the Jastrow correlated SD form (JSD). The last step, applied in this work only to some limited cases, is to release the lowest-rank condition of the geminal in Eq.~\ref{eq:QMC_wave function}, in order to introduce AGP correlations (JAGP).\cite{Casula2003}

The variational energy of the different $\Psi_{\rm VMC}$ flavors, namely JDFT, JSD, and JAGP, is computed by VMC. It is further projected towards the ground state energy by means of the DMC framework in the LRDMC scheme.\cite{Casula2005} Was the fermionic sign problem not present in the system, DMC would be exact. This is not the case except for a single H$_2$ molecule, whose LRDMC energy is exact because the spatial part of its ground state wave function is nodeless.\cite{Hongo2007} For the other cases analyzed here, i.e., the different supramolecular (H$_2$)$_2$ configurations, the fixed node approximation (FNA) has been used in our LRDMC calculations, leading to projected wave functions sharing the same nodes as the previously optimized $\Psi_{\rm VMC}$, called in this context trial wave function\cite{Pierleoni2008}. This approximation preserves the upper bound property of the LRDMC energies, which can therefore be systematically lowered by improving the quality of the nodes. For our system, we compared the DFT, SD, and AGP nodes in same selected cases, by studying their impact on the corresponding FNA-LRDMC energies, with the aim of assessing any residual bias coming from the FNA in our benchmark QMC results.

Other sources of bias in the LRDMC method are the energy dependence on the lattice space discretization of the Laplacian ($a_\textrm{latt}$) and the finite walker population bias. They can both be corrected systematically. For the former, we performed an accurate extrapolation of the dissociation energy as a function of $a_\textrm{latt}$. We found that a lattice space of $a_\textrm{latt}=0.125 a_0$, with $a_0$ the Bohr radius, is enough to provide energies whose discrepancy from extrapolated dissociation curves is within our statistical target error of 1 K (see Fig.~1 in the supplementary material (SM)\cite{SM}). The last bias is due to the finite walker population. For the LRDMC branching step, we used a fixed population algorithm with 256 walkers, yielding nearly unbiased energies. The residual bias, of the order of a fraction of a Kelvin, is removed by an energy extrapolation based on walkers reweighting.\cite{buonaura1998numerical} \\

Previous works on various systems have proven the very high accuracy of the LRDMC method.\cite{Marchi2009,Al-Hamdani2017,raghav2023toward} The same approach used here has already been applied on a related H$_4$ molecular complex made of four hydrogen atoms interacting in a square arrangement, showing that DMC 
outperforms standard quantum chemistry methods.\cite{gasperich2017h4,genovese2019assessing} We can therefore expect DMC to deliver benchmark quality results on (H$_2$)$_2$ dimers as well. To perform our QMC calculations, we have employed the \texttt{TurboRVB} package \cite{Nakano2020} throughout all the steps, from the generation of initial LDA orbitals using a built-in DFT driver to the final LRDMC simulations.

\subsection{\label{subsec:Part2B} Orbital-dependent functionals based on RPA}
Common approximate exchange-correlation (xc) functionals within DFT have a dependency on the density, the gradient of the density, and possibly the kinetic energy density.\cite{Perdew2001} More advanced functionals, such as those derived from many-body perturbation theory, usually incorporate a dependency on the Kohn-Sham (KS) orbitals,\cite{Kummel2008} which not only provides a better description of the atomic shell structure, but also substantially improves the quality of the xc energy and xc potential.\cite{jiang2005,Hellgren2010} 

The simplest orbital-dependent functionals are the hybrid functionals, which mix in a fraction of exact exchange into, e.g., a functional of the generalized gradient approximation (GGA). The PBE0 approximation, which is based on the PBE functional, includes 25\% of exact exchange.\cite{Adamo1999} However, since there is no strong justification for using 25\%, this fraction is adjustable. For this reason, we here instead refer to the generalized PBE0 functional, PBE0$\a$, where $\a$ corresponds to the percentage of exact exchange included.\cite{Perdew1996} With this notation, the standard PBE0 functional is written as PBE025. The xc energy of PBE0$\a$ reads:
\bea
E_{\rm xc}^{{\rm PBE0}\a} &=&- \frac{\tilde\a}{4}\int \gamma(\mathbf{r},\mathbf{r}') v(\mathbf{r}-\mathbf{r'}) \gamma(\mathbf{r'},\mathbf{r}) \, d\mathbf{r}d\mathbf{r'} \nonumber \\
&&+(1-\tilde\a)E_{\rm x}^{\rm PBE}+E_{\rm c}^{\rm PBE}.
\eea
Here, $\tilde\a=\a/100$, $\g$ is the first order spin-averaged reduced density matrix (assuming closed-shell systems), and $v$ is the Coulomb interaction.
The orbital dependence enters via $\g$, which contains a sum of the occupied KS orbitals. 

Since the KS orbitals have an implicit dependence on the density via the KS potential, the corresponding xc potential, $v_{\rm xc}=\d E_{\rm xc}/\d n$, is obtained by solving the optimized effective potential equation (OEP)\cite{Sharp1953,Talman1976}
\begin{equation}
\begin{aligned}
\displaystyle \int \chi_{\rm s} & (\mathbf{r}, \mathbf{r'}) v_{\rm x}(\mathbf{r'}) \, d\mathbf{r'} = \\
& - \int V_{\rm x}(\mathbf{r'},\mathbf{r''}) \Lambda_s(\mathbf{r}, \mathbf{r'}, \mathbf{r''}) \, d\mathbf{r'} \, d\mathbf{r''},
\end{aligned}
\label{eq:OEP_Equation}
\end{equation}
where $V_{\rm x}(\mathbf{r},\mathbf{r'})=-\frac{1}{2}\,v(\mathbf{r}-\mathbf{r'}) \, \gamma(\mathbf{r'},\mathbf{r})$. The functions $\chi_s$ and $\L_s$  are the independent-particle KS linear density, and linear density matrix response functions, respectively. The PBE xc contribution is added according to
\be
v^{\rm OEP\a}_{\rm xc}(\mathbf{r}) = \tilde\a v_{\rm x}(\mathbf{r}) +(1-\tilde\a)v_{\rm x}^{\rm PBE}(\mathbf{r}) +v_{\rm c}^{\rm PBE}(\mathbf{r}).
\label{oep_potential}
\ee
\begin{figure}[t]
\begin{center}
\includegraphics[scale=0.38,angle=0]{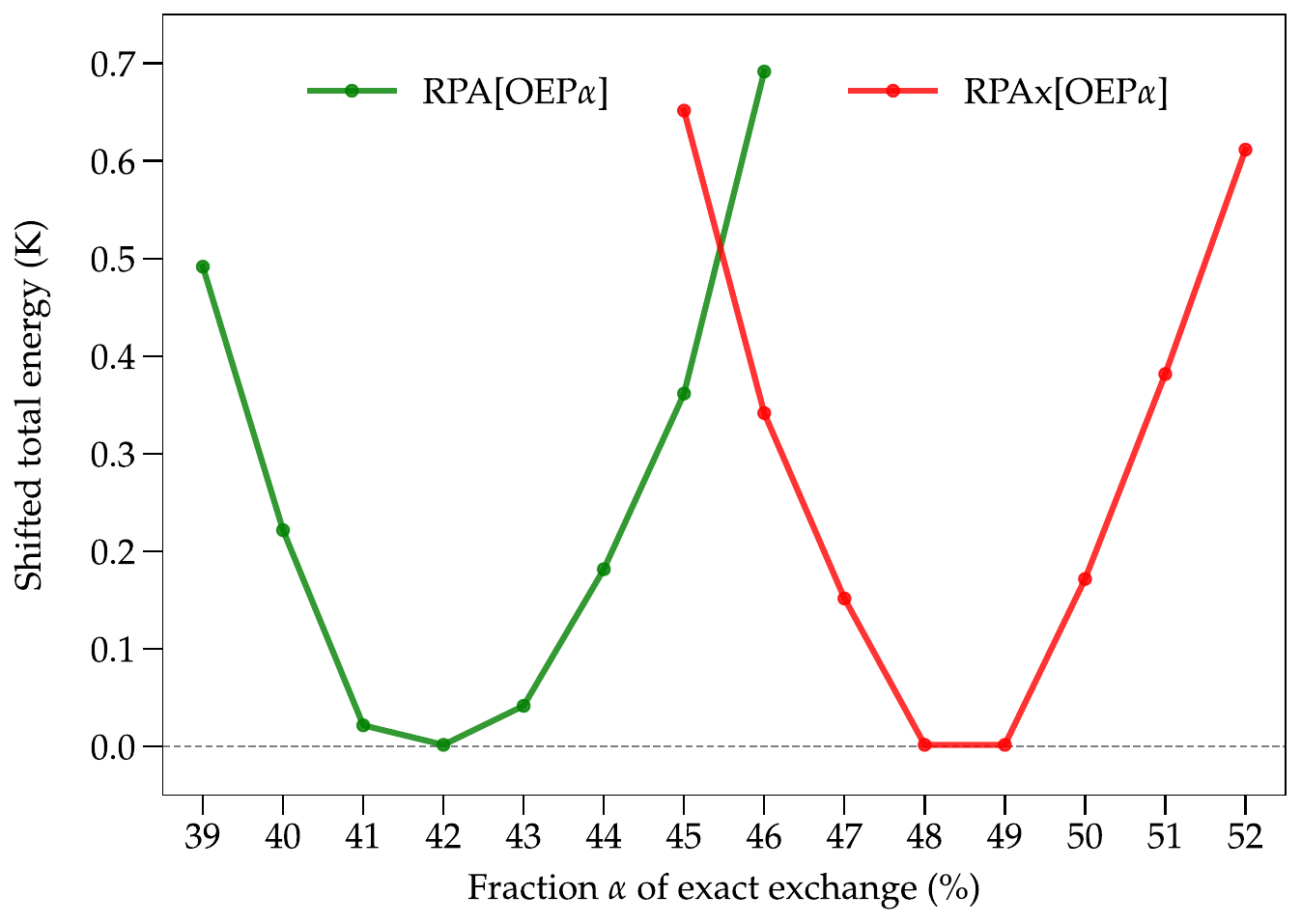}
\caption{\label{fig:Figure2_Main}Optimization of the proportion $\alpha$ of exact exchange included in OEP$\alpha$. RPA[OEP$\alpha$] and RPAx[OEP$\alpha$] data have been obtained on an isolated H$_2$ molecule at the equilibrium geometry predicted by RPA[OEP42] ($R_{\rm intra}$ = 0.742 \AA) and RPAx[OEP48] ($R_{\rm intra}$ = 0.741 \AA), respectively. The total energy is shifted to align the minimum of the curves with a zero reference line.}
\end{center}
\end{figure}

Hybrid functionals are sometimes implemented within the generalized KS scheme.\cite{Seidl1996} Instead of solving the OEP equation, the xc potential is allowed to be nonlocal as in the HF method. This has, in general, only a small impact on the density, but leads to a different eigenvalue spectrum.\cite{Baerends2013} In order to distinguish these two hybrid functional implementations, we, hereafter, use the notation PBE0$\a$ when referring to the generalized KS scheme, and OEP$\a$ when referring to the KS OEP scheme.
 
More advanced orbital-dependent correlation energy functionals can be generated through the ACFDT (adiabatic connection fluctuation dissipation theorem) formalism.\cite{langreth1975,langreth1977,Lein2000,Fuchs2002,Furche2005,ren2012random,eshuis2012electron,hesselmann2011} The ACFDT correlation energy is an exact expression for the KS correlation energy and is written as
\begin{equation}
\displaystyle E_{\rm c} = -\int \limits_{0}^{1} d\lambda \int \frac{d\omega}{2\pi} \mathrm{Tr} \Bigl \{v  \bigl [\chi_{\lambda} (i\omega) - \chi_{\rm s}(i\omega) \bigr ] \Bigr \},
\label{eq:RPA_correlation_energy}
\end{equation}
where $\chi_\l$ is the interacting linear density response function of a system described by the Hamiltonian 
\begin{equation}
\displaystyle \hat{H}[\lambda]= \hat{T} + \hat{V}_{\rm ext} + \lambda \hat{W} + \hat{V}_{\lambda}.
\label{eq:RPA_ACFDT}
\end{equation}
This fictitious Hamiltonian contains a parameter $\l$ that linearly scales the electron-electron interaction operator $\hat W$, and a potential operator $\hat V_{\l}$ that ensures the density to remain fixed to the physical density for every value of $\l$. This implies that $\hat V_{\l=1}$ is zero and that $\hat V_{\l=0}$ is equal to $\hat V_{\rm Hxc}$, the exact Hartree and xc potential operator. In Eq.~(\ref{eq:RPA_ACFDT}), $\hat T$ is the kinetic energy operator and $\hat V_{\rm ext}$ is the external potential operator. 

In order to generate approximate correlation energy functionals from Eq.~(\ref{eq:RPA_correlation_energy}), we need approximate expressions for $\chi_{\l}$. 
A convenient approach for generating such response functions is time-dependent density functional theory,\cite{tddftbook} which leads to a Dyson equation of the form
\begin{equation}
\displaystyle \chi_{\lambda} = \chi_{\rm s} + \chi_{\rm s} \Bigl [ \lambda v + f_{\rm xc}^{\lambda} \Bigr ] \chi_{\lambda} = \chi_{\rm s} + \chi_{\rm s} \Biggl [ \lambda v + \frac{\delta v_{\rm xc}^{\lambda}}{\delta n} \Biggr ] \chi_{\lambda}.
\label{eq:RPA_density_response_function}
\end{equation}
Approximations to $ \chi_{\lambda}$ thus originate from approximations to the linear response xc kernel $f_{\rm xc}$.\cite{Gross1985} Ignoring xc effects completely yields the RPA, or time-dependent Hartree, response function, which, through Eq.~(\ref{eq:RPA_correlation_energy}), defines the RPA correlation energy. The RPA is known to significantly improve the description of the xc energy. Contrary to most functionals in DFT, it provides a good description of molecules dissociating into open-shell fragments by capturing effects of static correlation.\cite{Fuchs2005,Hellgren2012} Furthermore, the vdW forces are seamlessly built in. The RPA can be further improved by including the exact-exchange kernel, $f_{\rm x}$, of the time-dependent OEP approach.\cite{tdOEP} Inserted into Eqs.~(\ref{eq:RPA_correlation_energy}) and (\ref{eq:RPA_density_response_function}) defines the RPAx approximation to the correlation energy. The RPAx has been shown to, e.g, improve the quantitative description of the vdW forces.\cite{Hellgren2010,hesselmann2010,Zhu2010,Bleiziffer2015,Hellgren2016,Hellgren2018,Hellgren2021} In this work, we have applied both RPA and RPAx, non-self-consistently.
\begin{table*}[ht!]
\caption{\label{tab:Table1_Main} Equilibrium properties of the five different (H$_2$)$_2$ configurations as predicted by LRDMC and CCSD(T).\cite{Patkowski2008,Lu2013}
}
\vspace{5pt}
\centering

\begin{tabular}{M{4.5cm} M{2cm} M{2cm} M{2cm} M{2cm} M{2cm} N}
    \toprule
    \multicolumn{1}{c}{} & \multicolumn{5}{c}{\textbf{Dissociation energy \textit{D}$_{\rm \textbf{e}}$ (K)}} \\
    \cline{2-6}
    \textbf{} & \textbf{T-shaped} & \textbf{Z-shaped} & \textbf{X-shaped} & \textbf{H-shaped} & \textbf{L-shaped} &\\
    \hline
    LRDMC & 53.43(70) & 50.46(41) & 26.46(39) & 19.71(54) & 15.14(55) &\\
    CCSD(T) {\footnotesize [Patkowski et al.]} \cite{Patkowski2008} & 55.04 & / & 25.14 & 18.23 & 13.48 &\\
    CCSD(T) {\footnotesize [Lu et al.]} \cite{Lu2013} & 52.1 & 48.1 & 24.3 & 17.6 & 14.0 &\\
    \toprule
\end{tabular}

\vspace{20pt}

\begin{tabular}{M{4.5cm} M{2cm} M{2cm} M{2cm} M{2cm} M{2cm} N}
    \toprule
    \multicolumn{1}{c}{} & \multicolumn{5}{c}{\textbf{Equilibrium intermolecular distance \textit{R}$_{\rm \textbf{eq}}$ (\AA)}} \\
    \cline{2-6}
    \textbf{} & \textbf{T-shaped} & \textbf{Z-shaped} & \textbf{X-shaped} & \textbf{H-shaped} & \textbf{L-shaped} &\\
    \hline
    LRDMC & 3.359(5) & 3.380(3) & 3.492(4) & 3.562(8) & 3.729(7) &\\
    CCSD(T) {\footnotesize [Patkowski et al.]} \cite{Patkowski2008} & 3.36 & / & 3.49 & 3.57 & 3.77 &\\
    CCSD(T) {\footnotesize [Lu et al.]} \cite{Lu2013} & 3.405 & 3.435 & 3.585 & 3.678 & 3.721 &\\
    \toprule
\end{tabular}

\end{table*}

For the DFT, HF, and PBE0$\a$ calculations, we have used the \texttt{PWscf} package of the plane-wave \texttt{Quantum ESPRESSO} (\texttt{QE}) code.\cite{Giannozzi2017} The self-consistent OEP$\a$, as well as the non-self-consistent RPA and RPAx calculations, were performed using an updated version of the \texttt{QE} \texttt{ACFDT} package.\cite{Nguyen2009,Nguyen2014,Colonna2014,Hellgren2021} For each approximation, we used a PBE optimized norm-conserving Vanderbilt pseudopotential for the hydrogen atom.\cite{Hamann2013} We converged energy differences with an error less than 1 K using a cell-size of 30 Bohr and a plane-wave cutoff of 80 Ry. For the non-self-consistent RPA and RPAx calculations, a cell-size of 28 Bohr and a plane-wave cutoff of 60 Ry were sufficient. In the \texttt{QE} \texttt{ACFDT} implementation, the RPA/RPAx correlation energy is expressed in terms of a generalized eigenvalue problem involving the independent-particle KS response function. The eigenvalue problem is solved at every frequency using iterative diagonalization and density functional perturbation theory. \cite{Nguyen2009,Nguyen2014,Colonna2014}. In our calculations, the frequency integral in Eq.~(\ref{eq:RPA_correlation_energy}) was converged with a grid of 16 frequency points,\cite{Hellgren2018} while the RPA and RPAx response functions (Eq.~(\ref{eq:RPA_density_response_function})) were converged with a total of 20 eigenvalues per electron (see Fig.~2 in the SM\cite{SM}).

PBE orbitals were used to evaluate the non-self-consistent RPA/RPAx total energy (indicated in square brackets as RPA[PBE] and RPAx[PBE]). In order to improve the description of the orbitals, we also used hybrid OEP$\a$ orbitals, obtained by solving the OEP equation (Eqs.~(\ref{eq:OEP_Equation}) and (\ref{oep_potential})).
To identify the appropriate fraction of exact exchange to use, we optimized the mixing parameter $\alpha$ on an isolated H$_2$ molecule by minimizing the total energy in RPA[OEP$\alpha$] and RPAx[OEP$\alpha$] as a function of $\a$ (see Fig.~\ref{fig:Figure2_Main} and Ref. \onlinecite{Hellgren2022}). This corresponds to a constrained optimization that should yield orbitals as close as possible to the self-consistent RPA/RPAx orbitals. 
An energy minimum located far above 25\% can be identified in Fig.~\ref{fig:Figure2_Main}. RPA predicts an optimized exchange fraction of 42\% and RPAx a slightly larger fraction, 48\%. These values remain similar on the (H$_2$)$_2$ dimer (see Fig.~3 in the SM\cite{SM}). In the following, we will use these two fractions in our RPA and RPAx calculations (RPA[OEP42] and RPAx[OEP48]) as a first step towards full self-consistency. 

\section{\label{sec:Part3} Potential energy surface of (H$_2$)$_{2}$}
For a complete description of the PES of the (H$_2$)$_2$ dimer, there are six degrees of freedom to consider:\cite{Patkowski2008} the two intramolecular distances $R^1_{\rm intra}$ and $R^2_{\rm intra}$, the intermolecular distance $R_{\rm inter}$, defined as the distance between the centers of mass of the two H$_2$ molecules, and three angles describing the relative orientation between the two H$_2$ molecules. In this work, we have focused on five different configurations, identified in previous works, and labelled as the T-, Z-, X-, H- and L-shaped dimers (see Fig.~\ref{fig:Figure1_Main}).\cite{Diep2000,Patkowski2008,Lu2013} $R_{\rm inter}$ was then varied while keeping $R^1_{\rm intra}=R^2_{\rm intra}=R_{\rm intra}$ fixed at the value of an isolated H$_2$ molecule. This so-called rigid rotor approximation has been found to be valid within the range of intermolecular distances considered.\cite{Ree1979} All five dimers exhibit a minimum along the $R_{\rm inter}$ potential energy curve, which we will call the equilibrium or bond distance $R_{\rm eq}$. We note, however, that it is only the T-shaped dimer that does not have imaginary vibrational frequencies in the harmonic approximation.\cite{Lu2013} We also note that due to the zero point vibrational energy, the molecule is estimated to exist in a mixture of configurations with a binding energy as small as 4-6 K, according to an analytical potential fit.\cite{Khan2020} 

\subsection{\label{subsec:Part3A}Dissociation energies}
We first determined, for every approximation and method, the energy and bond distance of the isolated H$_2$ molecule (see Table I in the SM\cite{SM}). The dissociation energy of the (H$_2$)$_2$ dimer was then determined as the difference between the total energy of the dimer at equilibrium distance $R_{\rm eq}$ and twice the total energy of an isolated H$_2$ molecule
\begin{equation}
D_{\rm e}= E_{\rm dimer}-2E_{\rm molecule }.   
\end{equation}
Accurate reference data was provided using the LRDMC method. In order to evaluate the influence of the quality of the trial wave function on the results, we have carried out LRDMC simulations by using trial wave functions with LDA nodes (JDFT) and with nodes optimized at the VMC level, obtained by relaxing the one-body orbitals of the JSD wave function (see Fig.~4 in the SM \cite{SM}). Calculations have been performed for intermolecular distances within the range 3.0-4.4 \AA~for all five dimers, with steps of 0.2 \AA. To reach the target statistical error, namely a standard deviation for the energy below 1 K, we generated a total number of LRDMC configurations per walker as large as 22.5M, yielded by a multi-walker algorithm involving 256 walkers.
On general grounds, improving the quality of the nodes leads to better results. We have analyzed the quality of the trial wave function by calculating its variance around the equilibrium geometry for every dimer. On average, we find a variance on the wave function six times smaller after the VMC optimization of the LDA orbitals, about 0.003 Ha$^2$ against 0.018 Ha$^2$. Nevertheless, both values are small, and as illustrated in Fig.~4 in the SM, \cite{SM} the change in the LRDMC dissociation energies after the nodes optimization is minor. The nodal bias is within statistical error bars in the stretched region of the dimers PES beyond 3.5 \AA, while it becomes more sizable in the compressed part of the dissociation curves, where it exceeds the statistical error for the majority of dimers geometries. Indeed, in the repulsive region JDFT-LRDMC gives steeper walls than JSD-LRDMC. Nevertheless, the overall quality of the LDA nodes on (H$_2$)$_2$ systems is already decent, particularly if one looks at equilibrium properties, where it yields results compatible with CCSD(T) calculations.  To assess the LRDMC results obtained within the fully optimized JSD trial wave function, we then tried to improve the JSD-based results even further by assuming the wave function to be expressed as JAGP instead of JSD, on some selected geometry points. We notice that the JAGP dissociation energies are within error bars from the JSD energies, with no significant improvement in the range of intermolecular distances studied here. This is in line with what stated in Ref.~\onlinecite{genovese2019assessing}.
The same observation can be made when monitoring the change in the variance of the wave function. While the optimization of the nodes at the JSD level has led to a decrease of the variance by more than 80\% from the LDA nodes, the improvement with the AGP is mainly around 20\% for a mild increase in computational burden, but at the cost of a much more delicate optimization procedure. Therefore, choosing the trial wave function to be a Jastrow correlated single Slater determinant already provides us with a very accurate description of the different hydrogen dimers within the LRDMC approach for the range of distances taken into account in this work. Unless otherwise specified, the benchmark LRDMC results presented hereafter will be obtained using a JSD trial wave function.
\begin{figure}[t]
\centering
\includegraphics[scale=0.62,angle=0]{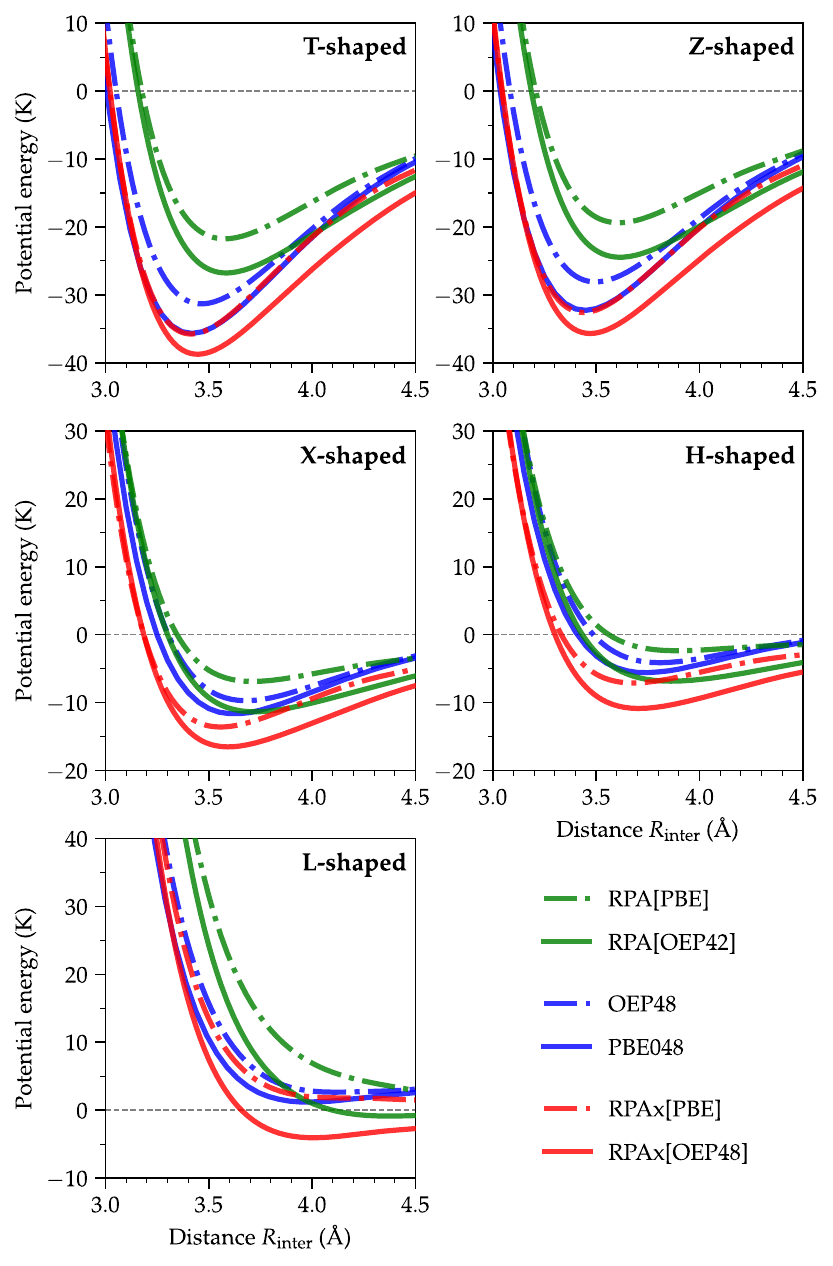}
\caption{Dissociation curves of the five different (H$_2$)$_2$ configurations (see Fig.~\ref{fig:Figure1_Main}). RPA[PBE] is compared to RPA[OEP42], OEP48 to PBE048, and RPAx[PBE] to RPAx[OEP48].}
\label{fig:Figure3_Main}
\end{figure}

\begin{figure}[t]
\begin{center}
\includegraphics[scale=0.62,angle=0]{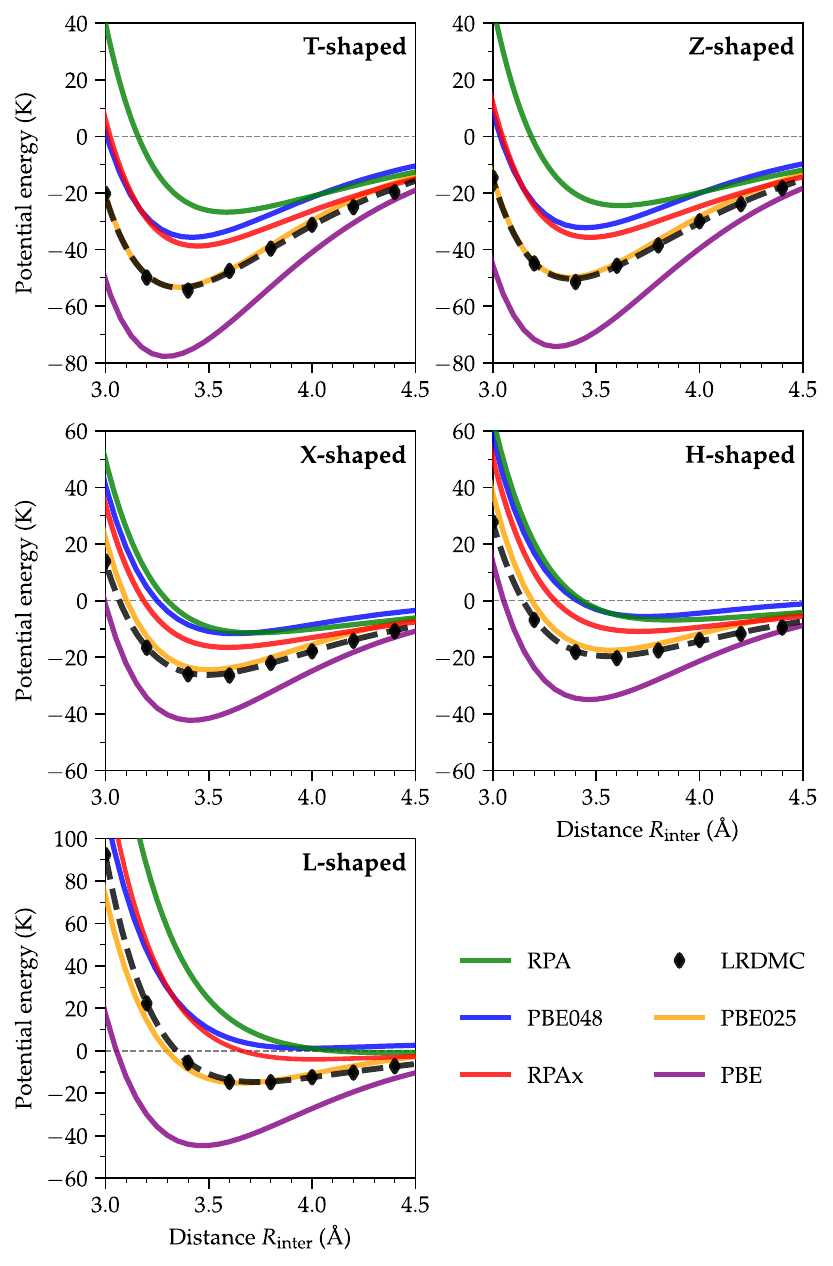}
\caption{\label{fig:Figure4_Main} Dissociation curves of the five different (H$_2$)$_2$ configurations (see Fig.~\ref{fig:Figure1_Main}). RPA[OEP42] and RPAx[OEP48] are compared to PBE048, PBE025, PBE and LRDMC. The fitted LRDMC curve is represented by black dashes.}
\end{center}
\end{figure}

\begin{figure}[t]
\begin{center}
\includegraphics[scale=0.62,angle=0]{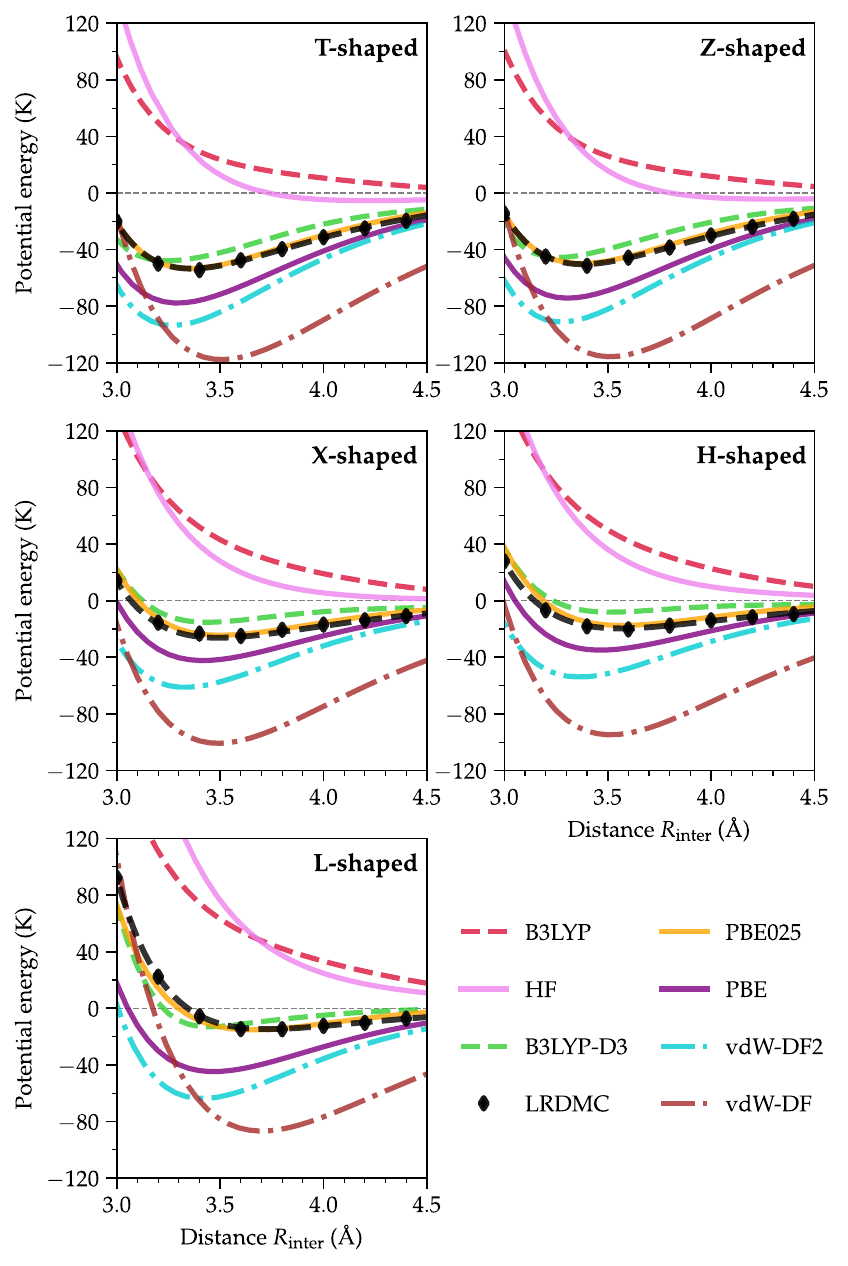}
\caption{\label{fig:Figure5_Main} Dissociation curves of the five different (H$_2$)$_2$ configurations (see Fig.~\ref{fig:Figure1_Main}). LRDMC, HF, and various approximations within DFT have been tested. The fitted LRDMC curve is represented by black dashes.}
\end{center}
\end{figure}
\begin{figure*}[hbt!]
\centering
\includegraphics[scale=0.60,angle=0]{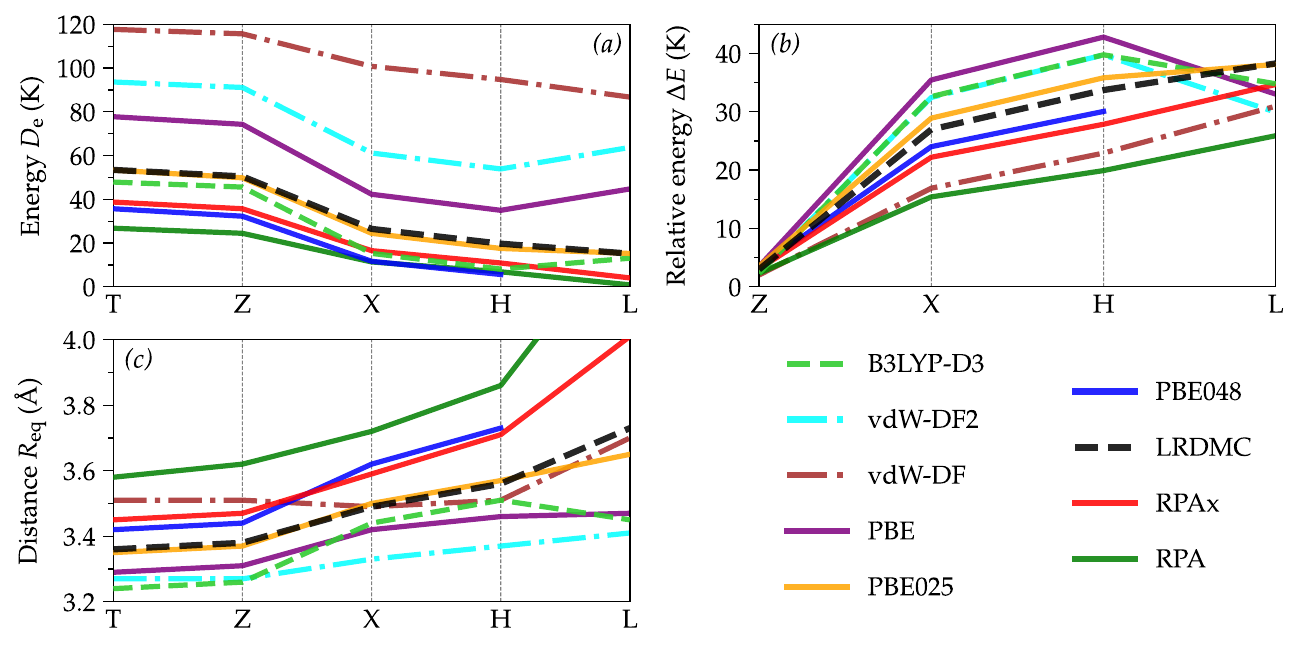}
\caption{Comparison between different approximation methods of \textit{(a)} the dissociation energy \textit{D}$_{\rm e}$ of each dimer, \textit{(b)} their relative energy $\Delta \rm{\textit{E}}$ with respect to the T configuration, and \textit{(c)} their intermolecular bond distance $R_{\rm eq}$.}
\label{fig:Figure6_Main}
\end{figure*}
To obtain smooth dissociation curves within LRDMC, we have fitted our data points with a Morse potential between 3.0 \AA~and 4.5 \AA, by also including the asymptotic limit that is two non interacting H$_2$ molecules. The bootstrap method has then been employed to reliably estimate the statistical error on the fitting parameters, given the error in the energy at each geometry point. From the generated curves, we have been able to determine the corresponding LRDMC dissociation energy $D_{\rm e}$ and equilibrium distance $R_{\rm eq}$ for each H$_2$-H$_2$ configuration (see Table~\ref{tab:Table1_Main}). These values are in a good overall agreement with CCSD(T) results. \cite{Patkowski2008,Lu2013} 

Despite the high accuracy displayed by LRDMC the computational cost is high.
It is, therefore, interesting to identify other methods that can offer a better accuracy-cost ratio.

Next, we generated data with RPA and RPAx. The potential energy curves of the five different dimers are presented in Fig.~\ref{fig:Figure3_Main}. In lack of a fully self-consistent implementation, RPA and RPAx have been evaluated with PBE and optimized OEP$\a$ orbitals (see Sec.~\ref{subsec:Part2B}). 
We have also included the corresponding RPAx-optimized hybrid functional, identified in Fig.~\ref{fig:Figure2_Main}, for comparison. The latter is evaluated  with orbitals generated by the nonlocal exchange potential (PBE048) and with orbitals generated by the local exchange potential (OEP48). We note that the use of a nonlocal or a local potential becomes equivalent for the total energy in the dissociation limit since, for a two-electron system, there is only one occupied orbital, turning the Fock exchange potential effectively local. This can be seen in  Fig.~\ref{fig:Figure3_Main}, where the two hybrid implementations start to merge as the dimer is stretched beyond a distance $R_{\rm{inter}}$ of about 4.2 \AA. Around equilibrium, the results differ by 1-5 K. 

For RPA and RPAx, we see significantly deeper potential energy wells when OEP$\a$ orbitals are employed, which suggests that self-consistency is of importance within these approximations. In RPA, the dissociation energy increases by more than 25\%. This is twice as much as in RPAx, and, hence, RPA appears even more sensitive to the choice of input orbitals as compared to RPAx. We also note that the improved input orbitals are crucial to observe a minimum in the potential energy curve of the L-shaped dimer. For simplicity, we will, from now on, denote RPA[OEP42] and RPAx[OEP48] as RPA and RPAx, respectively. 

A comparison with LRDMC is given in Fig.~\ref{fig:Figure4_Main}. RPA is able to recover around half of the dissociation energy. RPAx improves the results, capturing around 72\% of the dissociation energy. In the case of the (H$_2$)$_2$ dimer, the RPAx thus gives results similar to the MP2 method.\cite{Diep2000,Lu2013} Full self-consistency could improve results further but, for capturing the full dissociation energy, it seems necessary to include correlation effects in the xc kernel. 

In Fig.~\ref{fig:Figure4_Main}, RPA and RPAx are also compared to PBE025 and PBE, and in Fig.~\ref{fig:Figure5_Main}, an additional set of approximations (HF, B3LYP, B3LYP-D3, vdW-DF, vdW-DF2) are included. The complete set of data for the equilibrium properties are presented in Tables II, III and IV, in the SM.

The BLYP and PBE functionals are often used for studying solid hydrogen and it has been shown that BLYP performs slightly better 
than PBE.\cite{Morales2014} For the dimers, BLYP clearly gives an inferior description to PBE (see the SM\cite{SM} for BLYP results). While BLYP is unable to bind, PBE provides a reasonable, although overestimated, dissociation energy. A striking improvement is found when including a fraction of exact exchange. The fact that 25\% of exact exchange within PBE0 delivers such good results is surprising. Hybrid functionals are well known to encounter difficulties in describing weakly bonded systems due to missing vdW forces. In Refs.~\onlinecite{Patkowski2008} and \onlinecite{Lu2013}, an analysis in terms of energy decomposition was carried out in order to understand the nature of the (H$_2$)$_2$ intermolecular interaction around equilibrium. It was found that the exchange repulsion and the vdW component largely cancel such that the crucial interaction is electrostatic. The anisotropy of this region of the PES and the relative stability of the dimers are thus mainly determined by a quadrupole-quadrupole term. This could be the reason behind the very good performance of the PBE025 functional. Given that RPAx predicts the optimal fraction of exchange to be 48\%, it seems as if PBE025 can compensate for the missing vdW attractive force by underestimating the fraction of exchange. 

The B3LYP functional incorporates 20\% of exact exchange within the BLYP functional. Adding a vdW correction at the D3 level of theory produces rather accurate results. Although not as precise as PBE025, B3LYP-D3 appears to provide a better description of the various energy components as compared to PBE025. However, functionals that include the nonlocal vdW interaction, such as vdW-DF and vdW-DF2, strongly overestimate the dissociation energy. 
\begin{figure*}[hbt!]
\centering
\includegraphics[scale=0.70,angle=0]{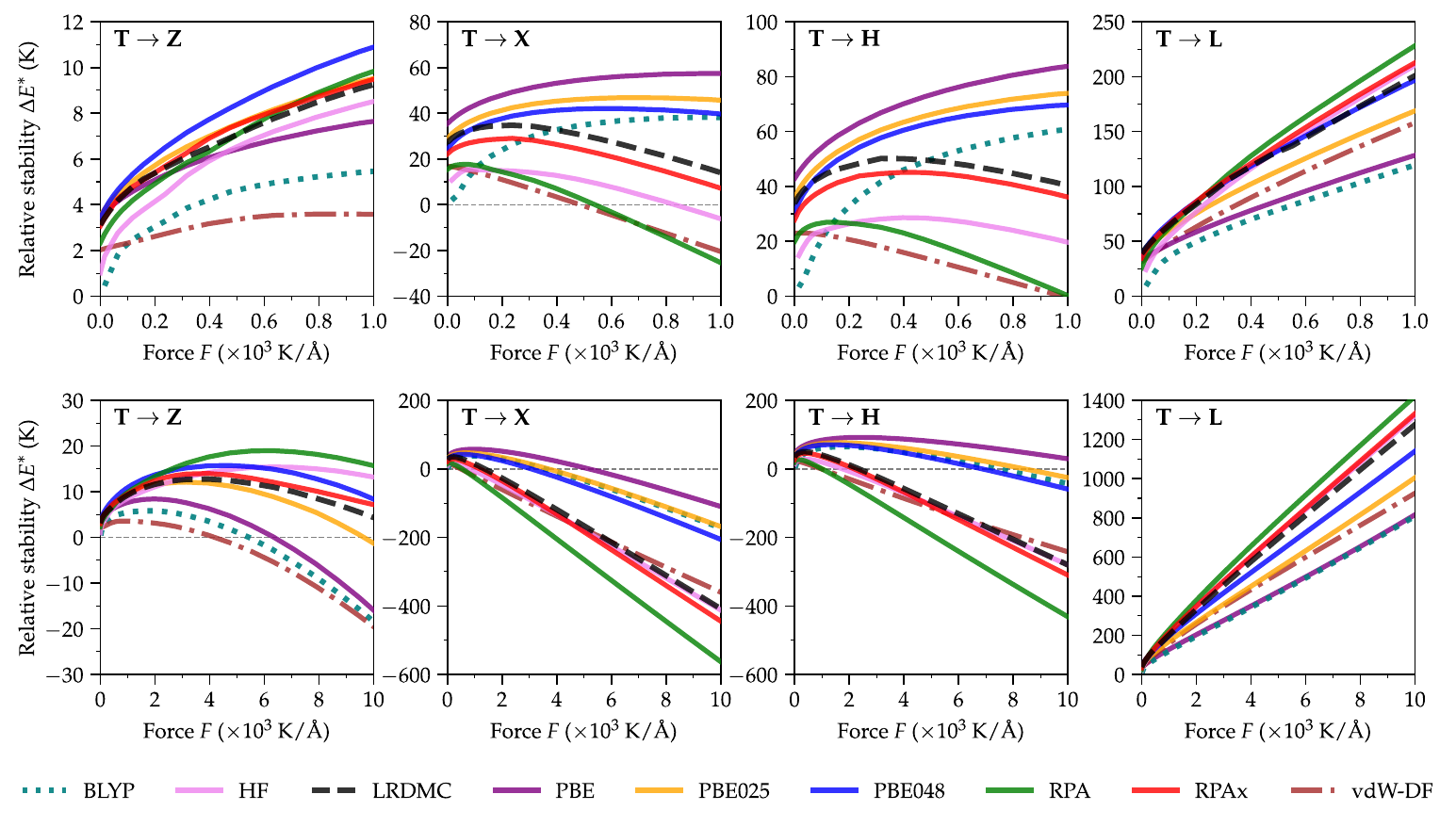}
\caption{\label{fig:Figure7_Main} Evolution of the relative stability, $\D E^*$, between the T-shaped dimer and the other four orientations (see Fig.~\ref{fig:Figure1_Main}) under the action of a weak compressive force (top panels), and under the action of a strong compressive force (bottom panels).}
\end{figure*}

\begin{figure}[t]
\begin{center}
\includegraphics[scale=0.62,angle=0]{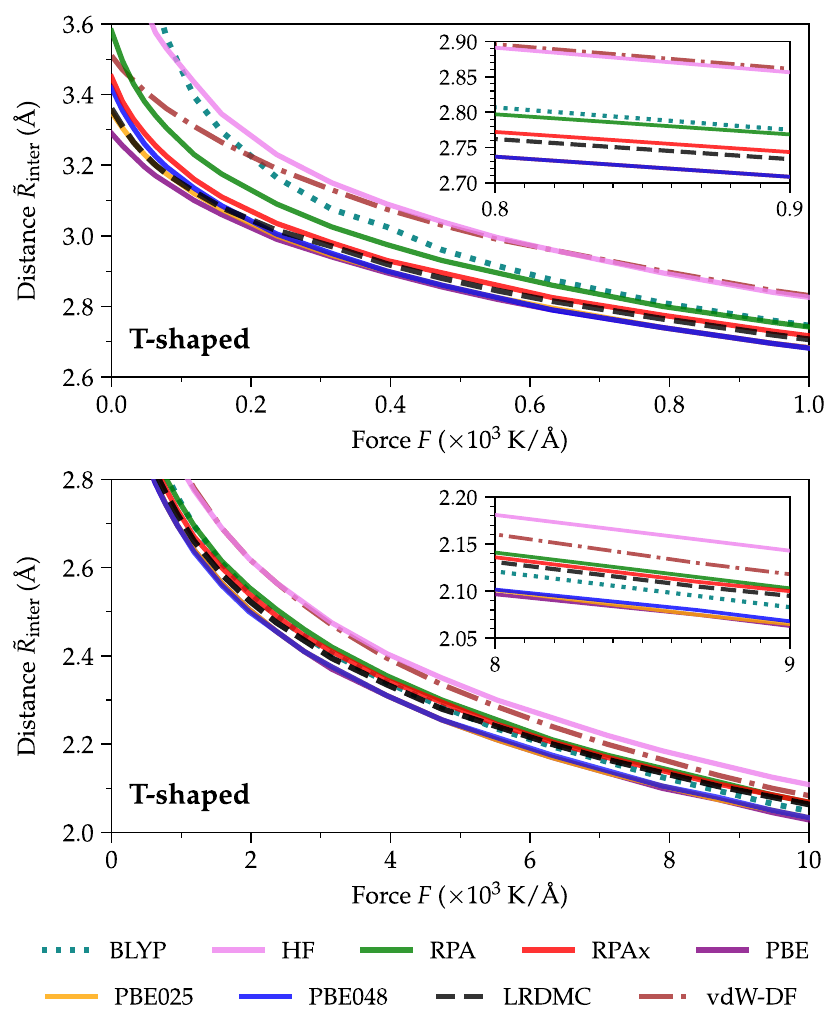}
\caption{\label{fig:Figure8_Main} Change in the intermolecular distance of the T-shaped dimer under the action of a weak compressive force (top panel), and under the action of a strong compressive force (bottom panel).}
\end{center}
\end{figure}
Results obtained for the dissociation energies, the
intermolecular bond distances and the relative energies, defined as the difference in energy of a dimer with respect to the T-configuration taken at equilibrium distances, are summarized in Fig.~\ref{fig:Figure6_Main} and in the SM (see, e.g., Fig. 5 in the SM). After PBE025, B3LYP-D3 performs the best in terms of a mean absolute error of the dissociation energies. 
RPAx and PBE048 also show good performance in terms of dissociation energies and give slightly better relative energies. In terms of bond distances, vdW-DF and PBE are better on the average, but have a less consistent performance over the various dimer configurations. Similarly, we note that PBE, vdW-DF2 and B3LYP-D3 all fail in determining the correct order of stability by overestimating the dissociation energy of the L-shaped dimer. It is thus worth mentioning that the only methods able to give a qualitatively correct picture of the relative energies are PBE025, vdW-DF, RPA, and RPAx. \\

In the next subsection, we will extend the analysis of the relative stability of the dimers by applying a compressive force along the intermolecular axis. In this way, the repulsive part of the PES is assessed. 

\subsection{\label{subsec:Part3B} Finite compression}
Molecular hydrogen condenses on a hcp lattice at low temperature and ambient pressure.\cite{Silvera1980,Mao1994,McMahon2012} Calculating the lattice energy, we expect the performance of the different approximations to be similar to that found for the isolated dimers. However, at finite pressures (100-300 GPa pressure range), intermolecular distances reduce to 1.5-2.5 \AA~and, hence, the description of the repulsive potential wall becomes more relevant.\cite{Ree1979,Burton1982,Roeggen1992_c} In order to gain insight into the performance of the different approximations on the high-pressure crystal phases, we here extend the analysis in Sec.~\ref{subsec:Part3A} to this region of the PES. 
 
Rather than looking at the energy as a function of the intermolecular distance, $E(R_{\rm inter})$, we compare the potential energy curves at a fixed force applied along the intermolecular axis, $F=-\partial E/\partial R_{\rm inter}$. To convert $E(R_{\rm inter})$ to a function of the force, we perform a Legendre transformation  
\be
E^*(F)=E(\tilde{R}_{\rm inter})+\tilde R_{\rm inter}F,
\label{enthalpy}
\ee
where $\tilde{R}_{\rm inter}$ and $F$ are coupled through the minimization of $E(R_{\rm inter}) + R_{\rm inter}F$ with respect to $R_{\rm inter}$, at fixed $F$. The quantity $E^*$ is similar to the enthalpy as defined for extended systems under pressure. We will, therefore,  interpret the difference, $\D E^*$, between different dimer configurations, as a measure of the relative stability. 
The study of $\D E^*$ is a natural extension of the analysis of the relative energies in Sec.~\ref{subsec:Part3A}, which corresponds to setting $F=0$ in Eq.~(\ref{enthalpy}).

In Fig.~\ref{fig:Figure7_Main}, we present the results obtained for $\D E^*$ as a function of the applied force, comparing the different dimers with respect to the T-configuration, which is the most stable orientation at equilibrium. We use the same methods and approximations considered at zero compression. The scan along $R_{\rm inter}$ was done with a step size of 0.05 \AA. To generate smooth curves, the step size was decreased to 0.005 \AA~by fitting the data with two different polynomial functions: the first close to the minimum identified at zero compression, the second in the repulsive region down to 2.0 \AA. These two fits have then been merged at 3.0 \AA. A finite compression is expected to change the intramolecular distances $R^1_{\rm intra}$ and $R^2_{\rm intra}$ as well, if they were allowed to relax. However, in the range of forces considered here this change is less than 1\%.\cite{Ree1979} We, therefore, focus only on the relative orientations, and keep the intramolecular distance of the two hydrogen molecules fixed at $R_{\rm intra}$ optimized for the isolated H$_2$ molecule (see Table 1 in the SM \cite{SM}). 

To obtain a reference data set, calculations were first performed with the LRDMC method. To save computational power, the missing data points along $R_{\rm inter}$ between 2.0 \AA~and 3.0 \AA~were acquired with a step size of 0.2 \AA. They were then fitted with a Morse potential using the bootstrap technique to take statistical errors into account. Again, separate fits were done around equilibrium and in the repulsive wall. They were then merged at 3.0 \AA~to obtain a set of data with a decreased step size of 0.005 \AA. The fits capture well the statistical error associated to every data point calculated with LRDMC, thus validating the fitting procedure (see Fig. 6 of the SM \cite{SM}).

We also compared the LRDMC results with the results obtained by Ree et al. in Ref.~\onlinecite{Ree1979} that used CISD (Configuration Interaction with Single and Double excitations), and with those obtained by R$\o$eggen et al. in Refs.~\onlinecite{Roeggen1992_a,Roeggen1992_b}. The latter results were obtained with EXRHF3, a geminal-based model whose expansion is truncated at the double-pair correlation level, which should approach a full CI result in the (H$_2$)$_2$ system in the complete basis set limit (see Fig. 6 of the SM \cite{SM}). The good agreement found, especially with EXRHF3, confirms that our LRDMC results are of high quality and can be used as a benchmark in the compressed regime.

In the top panels of Fig.~\ref{fig:Figure7_Main}, the force takes values up to 10$^3$ K/\AA, and in the lower panels up to 10$^4$ K/\AA. This corresponds to intermolecular distances of about 3.6-2.6 \AA~and 3.6-2.0 \AA, respectively (see Fig.~\ref{fig:Figure8_Main}). Compared to the results at zero compression, a different picture now emerges. Most strikingly is the failure of the PBE025 functional, which quickly loses most of its agreement with LRDMC. A similar result is found for B3LYP-D3 (not shown in the Figure).
Only the RPAx is able to keep a consistent fairly good agreement with LRDMC over the whole range of forces, including the correct prediction of $\tilde{R}_{\rm inter}$ (see Fig.~\ref{fig:Figure8_Main}). 
The RPA, which does not include the exchange kernel, also performs relatively well at finite compression, especially under large compressive forces. On the other hand, most of the DFT functionals tested present significant disagreements. Clearly, the inclusion of exact exchange improves the performance of most functionals. The hybrids PBE025, PBE048 and B3LYP (not shown in the Figure) give better results than PBE and BLYP but, contrary to the picture at zero compression, PBE048 is now systematically better than PBE025. We expect that the vdW contribution to the energy is less relevant for describing the repulsive potential wall. Hence, a functional that relies on a compensation between the exchange repulsion and the vdW interaction may lose its performance under compression. This could be the reason why PBE025 now fails, for example. Overall, including a large fraction of exact exchange seems essential. Even HF performs well under a large compression. On the other hand, looking at the intermolecular distance in Fig.~\ref{fig:Figure8_Main}, we see that HF predicts too large values for $\tilde R_{\rm inter}$ at fixed force. Correlation is thus not negligible in this regime.

Let us now study how the difference, $\D E^*$, or the relative stability, of the different dimers evolves with applied force. It is perhaps not surprising to see that the L configuration is the hardest to compress (see Fig. 8 in the SM\cite{SM}). The H- and X-shaped dimers, both with the intramolecular axes in two different parallel-shifted planes, are the easiest to compress at large applied force. Eventually also the Z-shaped dimer becomes easier to compress than the T-shaped dimer. The configurational preference thus changes with compression. 
Within the set (T, Z, X, H), the T-dimer eventually becomes the least preferred. It is thus clearly difficult for approximate methods to accurately capture this evolution. The simpler functionals in DFT, such as PBE and BLYP, tend to favor the Z-dimer compared to all the others configurations, already at small compression. They also strongly underestimate the repulsive wall in the L-shaped dimer with respect to the T-dimer. The same functionals overstabilize the T configuration with respect to the X- and H-shaped dimers. Similar trends are seen for PBE025, but a large improvement in the L-shaped dimer is found with PBE048. Similarly to equilibrium, the vdW-DF functional gives an overall better description of the relative stability, in particular for the H- and X-dimers. However, it still favors the Z-dimer with respect to the T-dimer at small compression, and underestimates the repulsive wall in the L configuration.

In Ref.~\onlinecite{riffet2017}, it was shown that the main features of the various phase transitions in the solids can be simulated with a small cluster. In Ref. \onlinecite{Hellgren2022}, RPAx, RPA and PBE048 were identified among the best performing approximations for the study of high-pressure molecular solid hydrogen. This hierarchy of methods agrees well with our study of the dimers under compression. On the other hand, it is known that many-body effects tend to weaken the potential wall in the solid.\cite{Ree1979} Indeed, while HF works reasonably well on the compressed dimers, it strongly overestimates the effect of exchange in solids.\cite{liao2019comparative} 

Most of the model structures for the molecular solids have been generated with the PBE functional.\cite{Pickard2007} Looking at these structures, we find a predominance of tilted pairs of H$_2$ molecules, similar to the Z  configuration but with an angle different from 45°. Also H-dimers are relatively common in these structures. Of course, some of the possible molecular configurations could be less likely to appear due to stacking constraints. Nevertheless, given the rather large qualitative errors of the PBE functional in describing the relative stability of the various dimer configurations, it is not unlikely that new stable structures could be found using a more accurate functional.
 
\section{\label{sec:Part4} Conclusions}
An accurate description of the potential energy surface of the (H$_2$)$_2$ dimer represents a challenging problem for approximate many-electron methods. Capturing the delicate balance between exchange, vdW, and electrostatic forces, both at equilibrium and at finite compression, requires meV accuracy on each component. 

In this work, we have carried out accurate LRDMC calculations and assessed a number of functionals within DFT, including the advanced orbital-dependent RPA and RPAx functionals. Our results confirm the difficulties in capturing the PES of the (H$_2$)$_2$ dimer. Among the vdW functionals tested, only B3LYP-D3 produces accurate results around the potential well minima. The RPA underbinds which is most likely due to the well known problem of underestimated vdW forces. Including the exchange response kernel with RPAx significantly improves the well depths, but to capture the remaining 25\% of the dissociation energy a response kernel that includes correlation is necessary.  

A rather surprising result is the excellent performance of PBE025 (without vdW correction) around equilibrium. This could be due to a cancellation of errors, which, in this case, would be between missing vdW forces and an underestimation of the exchange repulsion. Indeed, at finite compression, the vdW component of the energy becomes less relevant in favor of the HF exchange, and PBE025, like most other functionals within DFT, cannot adapt and lose accuracy. Among the methods tested, only RPAx is able to give a consistent and satisfactory performance along the entire potential energy curve.

We also analyzed the performance of the different methods on the (H$_2$)$_2$ dimers with respect to their performance on the molecular solid phases at finite pressures. We have seen that the hierarchy of methods for the dimers at finite compression is similar to what is found for the high-pressure solids. On the other hand, while PBE0-based hybrid functionals show excellent performance on the solids as long as the fraction of exchange is properly tuned, on the dimers they are not sufficient and become rather inaccurate for certain dimer configurations. The reason for this difference could be that the structures for the solids have been generated with the PBE functional, which may favor orientations of the H$_2$ molecules better described by this type of functionals. Our analysis shows that the configurational preference is strongly functional dependent. 
Also, many-body effects and changes in the intramolecular distances are expected to affect the relative stability of the different (H$_2$)$_2$ orientations in the condensed phases, making the analysis on the dimers not directly transferable to the solids. Nevertheless, our analysis provides an indication of possible limitations in the performance of commonly applied approximate functionals. 

\section*{Supplementary Material}

\noindent The supplementary material provides tables containing the data for dissociation energies, bond distances, and relative energies with respect to the T-shaped dimer for the methods tested in this work. It also contains figures presenting the computational details of the LRDMC, RPA and RPAx calculations, as well as figures comparing LRDMC results with available data from the literature.

\section*{Data availability}

\noindent The data that support the findings of this study are openly available in Zenodo at\\ https://zenodo.org/records/13890532 \\ (reference number [DOI: 10.5281/zenodo.13890531]).

\acknowledgements
\noindent The work was partially performed using HPC resources from GENCI-TGCC/CINES/IDRIS (Grant No. A0150914650). M.C. acknowledges resources provided through the GENCI project number A0150906493 and through the TGCC special session (project number SS010915389).

\providecommand{\noopsort}[1]{}\providecommand{\singleletter}[1]{#1}

\widetext

\cleardoublepage 
\setcounter{figure}{0}    
{\noindent \LARGE \textbf{Supplementary Material}}

\vspace{10pt}

\begin{figure}[ht!]
\begin{center}
\includegraphics[scale=0.5,angle=0]{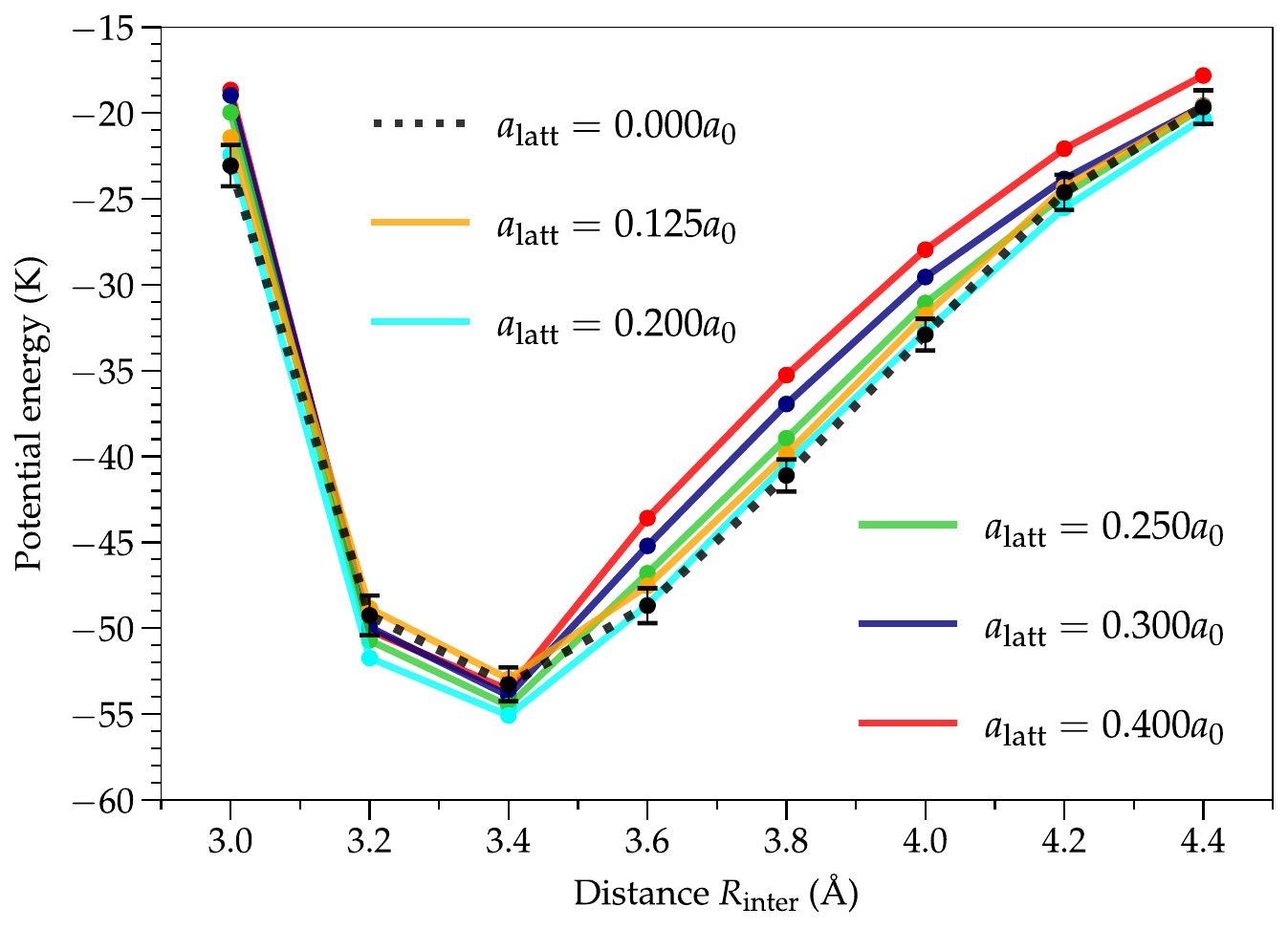}
\caption{\label{fig:Figure1_SM} Optimization of the lattice spacing parameter, $a_\textrm{latt}$, used in LRDMC calculations. The data at zero lattice spacing (black dotted curve) are estimated on the T-shaped dimer using the bootstrap method, by extrapolating the energy as an even polynomial function of $a_\textrm{latt}$, $f=f(a_\textrm{latt})$, with a varying part dependent on the power 2 and power 4 terms: $f(a_\textrm{latt}) = E_0 + c_2 a_\textrm{latt}^{2} + c_4 a_\textrm{latt}^{4}$. The estimated standard deviation on the extrapolated energy $E_0$ of each point of this fit is indicated in black square brackets.
}
\end{center}
\end{figure}

\newpage

\begin{figure}[ht!]
\begin{center}
\includegraphics[scale=0.5,angle=0]{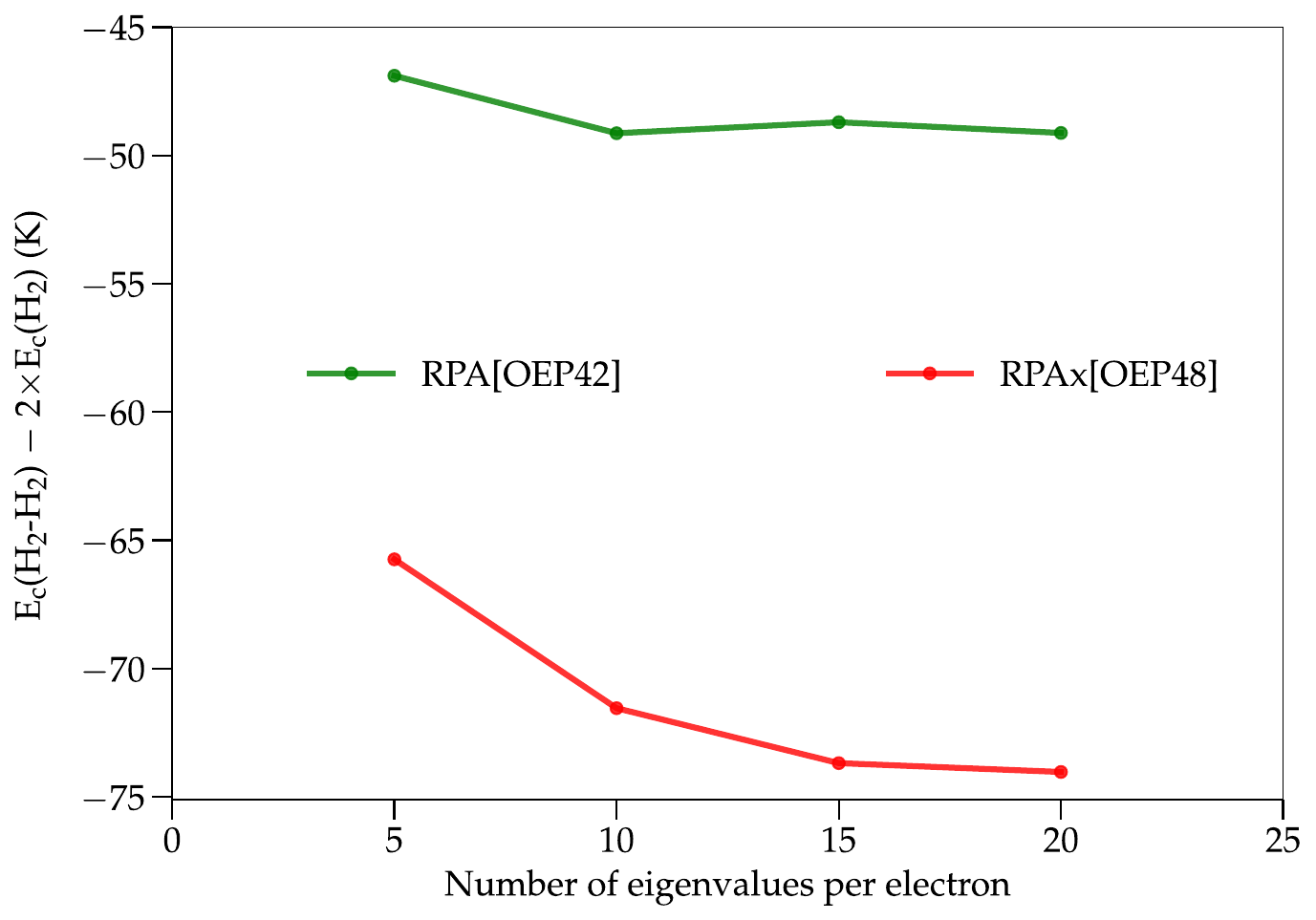}
\caption{\label{fig:Figure2_SM} Optimization of the number of eigenvalues used per electron to calculate the RPA/RPAx correlation energy E$_{\rm{c}}$. Data have been obtained on an isolated H$_2$ molecule and an isolated T-shaped dimer considered at the equilibrium geometry predicted by the RPA[OEP42] ($R_{\rm intra}$ = 0.742 \AA, $R_{\rm inter}$ = 3.58 \AA) and the RPAx[OEP48] ($R_{\rm intra}$ = 0.741 \AA, $R_{\rm inter}$ = 3.45 \AA) methods. These non-self-consistent RPA/RPAx calculations have been initialized using OEP-PBE0$\a$ orbitals (abbreviated OEP$\a$), where $\a$ is, in \%, the proportion of exact exchange included.}
\end{center}
\end{figure}

\begin{figure}[ht!]
\begin{center}
\includegraphics[scale=0.5,angle=0]{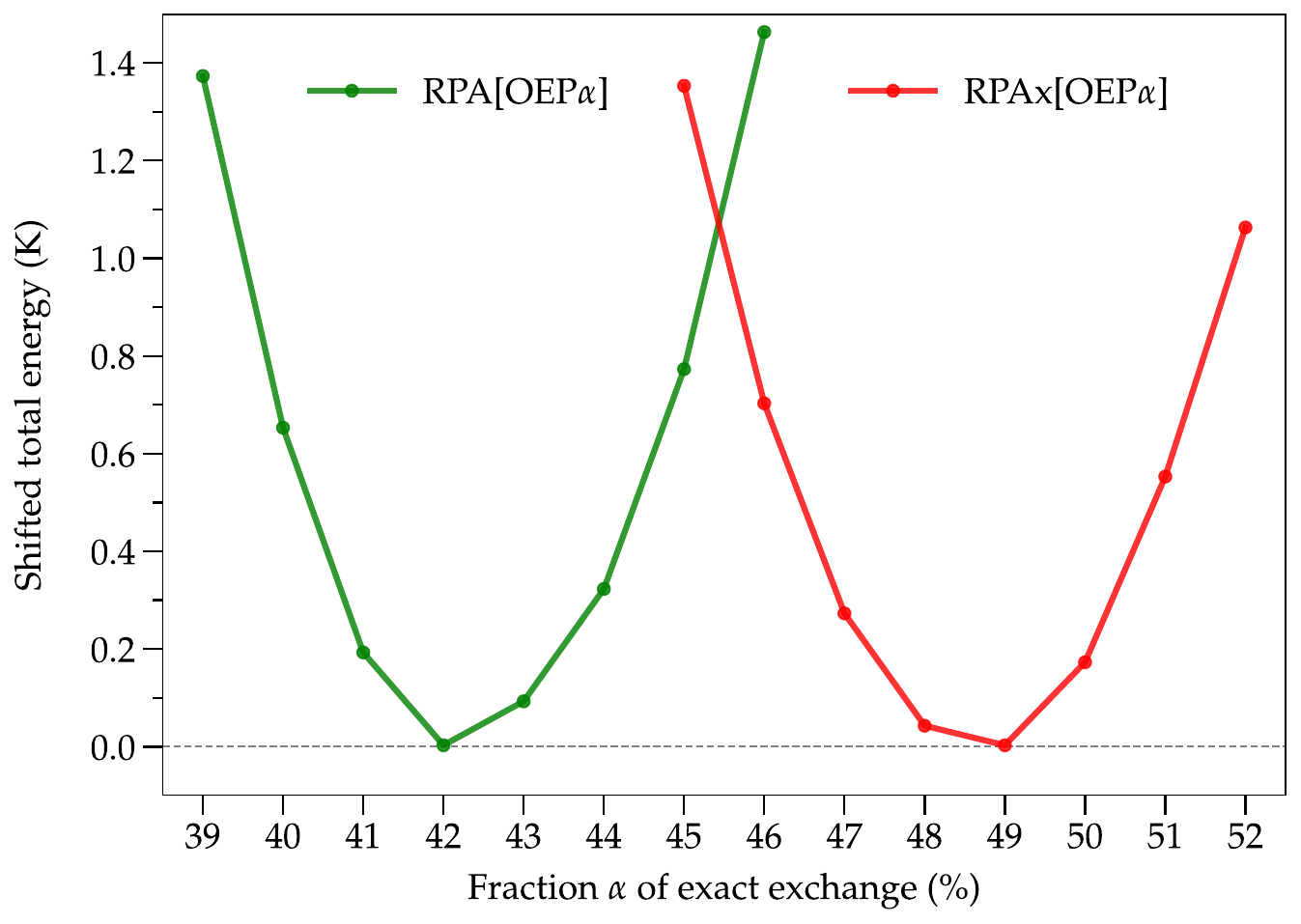}
\caption{\label{fig:Figure3_SM} Optimization of the proportion $\alpha$ of exact exchange included in OEP-PBE0$\alpha$ (abbreviated OEP$\alpha$). RPA[OEP$\alpha$] and RPAx[OEP$\alpha$] data have been obtained on an isolated L-shaped dimer considered at the equilibrium geometry predicted by RPA[OEP42] ($R_{\rm inter}$ = 4.37 \AA) and RPAx[OEP48] ($R_{\rm inter}$ = 4.01 \AA), respectively. The total energy is shifted to align the minimum of the curves with a zero reference line.}
\end{center}
\end{figure}

\newpage

\begin{figure}[ht!]
\begin{center}
\includegraphics[scale=0.74,angle=0]{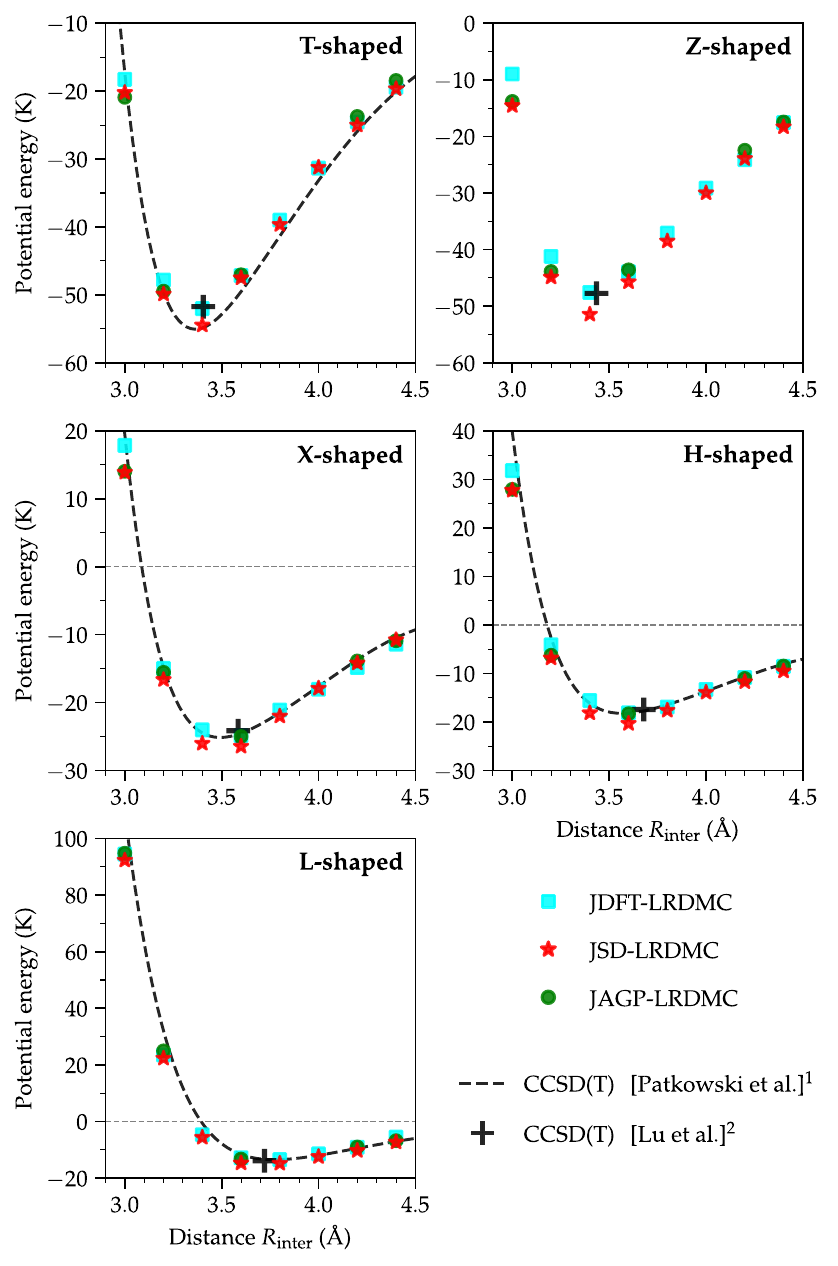}
\caption{\label{fig:Figure4_SM} Dissociation curves of five different (H$_2$)$_2$ configurations as predicted by the LRDMC within the fixed-node approximation. The trial wave function used is either of JDFT (cyan squares), JSD (red stars) or JAGP (green circles) type, the last two with nodes optimized at the VMC level. Calculations with the multireference JAGP wave function have been performed only on a subset of geometries covering the three main regions of the dissociation curve, namely the repulsive wall, the energy minimum, and the asymptotic regime. CCSD(T) reference data are indicated for the potential energy curve$^{1}$ and the position of the minimum$^{1,2}$ (black color).}
\end{center}
\end{figure}

\newpage

\begin{table}[ht!]
\caption{\label{tab:Table1_SM} Equilibrium geometry $R_{\rm intra}$ of the isolated H$_2$ molecule as predicted by several approximation methods. \\
}
\centering
\vspace{10pt}
\begin{tabular}{M{3.5cm} M{3.5cm} N}
    \toprule
    \textbf{} & \textbf{Distance $R_{\rm \textbf{intra}}$ (\AA)} \\
    \hline
    Xa & 0.781\\
    PZ & 0.765\\
    PBEsol & 0.757\\
    BP & 0.749\\
    PBE & 0.749\\
    PW91 & 0.747\\
    revPBE & 0.746\\
    BLYP & 0.745\\
    BLYP-D3 & 0.745\\
    PBE025 & 0.743\\
    RPA[PBE] & 0.742\\
    RPA[OEP42] & 0.742\\
    B3LYP & 0.741\\
    B3LYP-D3 & 0.741\\
    LRDMC & 0.741\\
    RPAx[OEP48] & 0.741\\
    RPAx[PBE] & 0.741\\
    X3LYP & 0.741\\
    OEP48 & 0.739\\
    PBE048 & 0.739\\
    vdW-DF & 0.738\\
    vdW-DF2 & 0.734\\
    HF & 0.732\\
    \toprule
\end{tabular}

\end{table}

\newpage

\begin{table}[ht!]
\caption{\label{tab:Table2_SM} Dissociation energy $D_{\rm e}$ of five different (H$_2$)$_2$ configurations according to several approximation methods. Data are reported only for the dimers predicted to be stable.}
\vspace{10pt}
\centering
\begin{tabular}{M{3.5cm} M{2cm} M{2cm} M{2cm} M{2cm} M{2cm} N}
    \toprule
    \multicolumn{1}{c}{} & \multicolumn{5}{c}{\textbf{Dissociation energy \textit{D}$_{\rm \textbf{e}}$ (K)}} \\
    \cline{2-6}
    \textbf{} & \textbf{T-shaped} & \textbf{Z-shaped} & \textbf{X-shaped} & \textbf{H-shaped} & \textbf{L-shaped} &\\
    \hline
    PZ & 324.19 & 314.55 & 176.42 & 157.20 & 321.74 &\\
    Xa & 195.57 & 188.52 & 99.06  & 85.14  & 174.50 &\\
    PW91 & 149.93 & 145.97 & 112.47 & 104.31 & 114.16 &\\
    vdW-DF & 117.67  & 115.64 & 100.77 & 94.75  & 86.71 &\\
    vdW-DF2 & 93.62  & 91.12  & 61.16  & 53.87  & 63.70 &\\
    PBE & 77.75      & 74.26  & 42.29  & 34.95  & 44.73 &\\
    PBEsol & 67.45   & 63.16  & 28.54  & 21.44  & 31.40 &\\
    LRDMC & 53.43    & 50.46  & 26.46  & 19.71  & 15.14 &\\
    PBE025 & 53.35   & 49.81  & 24.44  & 17.52  & 15.23 &\\
    B3LYP-D3 & 47.87 & 45.57  & 15.27  & 8.10   & 13.07 &\\
    RPAx[OEP48]&38.73& 35.67  & 16.52  & 10.87  & 4.05 &\\
    BLYP-D3 & 38.21  & 32.37  & 2.08   & 1.12   & 6.49 &\\
    RPAx[PBE]& 35.78 & 32.59  & 13.62  & 7.11   & /    &\\
    PBE048 & 35.67   & 32.26  & 11.65  & 5.61   & /    &\\
    OEP48 & 31.33    & 28.12  & 9.74   & 4.11   & /    &\\
    RPA[OEP42]&26.77 & 24.45  & 11.36  & 6.85   & 0.87 &\\
    revPBE & 26.71   & 25.23  & 14.51  & 11.01  & 7.93 &\\
    RPA[PBE] & 21.78 & 19.39  & 6.88   & 2.38   & /     &\\
    X3LYP & 10.69    & 8.00   & / & /  & / &\\
    HF & 5.34        & 4.33   & / & /  & / &\\
    B3LYP & 0.93     & 0.80   & / & /  & / &\\
    BP & 0.79        & / & /  & / & /  &\\
    BLYP & 0.71 &  / & / & /  & / &\\
    \toprule

\end{tabular}

\end{table}

\newpage

\begin{table}[ht!]
\caption{\label{tab:Table3_SM} Equilibrium geometry $R_{\rm eq}$ of five different (H$_2$)$_2$ configurations according to several approximation methods. Data are reported only for the dimers predicted to be stable.}
\vspace{10pt}
\centering
\begin{tabular}{M{3.5cm} M{2cm} M{2cm} M{2cm} M{2cm} M{2cm} N}
    \toprule
    \multicolumn{1}{c}{} & \multicolumn{5}{c}{\textbf{Equilibrium intermolecular distance $R_{\rm \textbf{eq}}$ (\AA)}} \\
    \cline{2-6}
    \textbf{} & \textbf{T-shaped} & \textbf{Z-shaped} & \textbf{X-shaped} & \textbf{H-shaped} & \textbf{L-shaped} \\
    \hline
    PZ & 2.76 & 2.76 & 2.87 & 2.90 & 2.80\\
    Xa & 2.96 & 2.97 & 3.09 & 3.13 & 3.02\\
    PBEsol & 3.22 & 3.24 & 3.47 & 3.55 & 3.39\\
    BLYP-D3 & 3.22 & 3.25 & 5.23 & 5.65 & 3.48 &\\
    B3LYP-D3 & 3.24 & 3.26 & 3.44 & 3.51 & 3.45 &\\
    PW91 & 3.25 & 3.26 & 3.35 & 3.35 & 3.42\\
    vdW-DF2 & 3.27 & 3.27 & 3.33 & 3.37 & 3.41\\
    PBE & 3.29 & 3.31 & 3.42 & 3.46 & 3.47\\
    PBE025 & 3.35 & 3.37 & 3.50 & 3.57 & 3.65\\
    LRDMC & 3.36 & 3.38 & 3.49 & 3.56 & 3.73\\
    RPAx[PBE] & 3.41 & 3.44 & 3.56 & 3.69 & /\\
    PBE048 & 3.42 & 3.44 & 3.62 & 3.73 & /\\
    X3LYP & 3.45 & 3.48 & / & / & /\\
    RPAx[OEP48] & 3.45 & 3.47 & 3.59 & 3.71 & 4.01\\
    OEP48 & 3.47 & 3.50 & 3.69 & 3.82 & /\\
    vdW-DF & 3.51 & 3.51 & 3.49 & 3.51 & 3.70\\
    RPA[PBE] & 3.56 & 3.61 & 3.72 & 3.86 & /\\
    RPA[OEP42] & 3.58 & 3.62 & 3.72 & 3.86 & 4.37\\
    revPBE & 3.86 & 3.88 & 3.95 & 4.03 & 4.23\\
    HF & 4.22 & 4.30 & / & / & /\\
    B3LYP & 5.83 & 6.04 & / & / & /\\
    BLYP & 6.24 & / & / & / & /\\
    BP & 6.26 & / & / & / & /\\
    \toprule
\end{tabular}

\end{table}

\newpage

\begin{table}[ht!]
\caption{\label{tab:Table4_SM} Relative energy $\D E$ with respect to the T configuration of the different dimers according to several approximation methods, at equilibrium geometry. Data are reported only for the dimers predicted to be stable.}
\vspace{10pt}
\centering
\begin{tabular}{M{3.5cm} M{2cm} M{2cm} M{2cm} M{2cm} N}
    \toprule
    \multicolumn{1}{c}{} & \multicolumn{4}{c}{\textbf{Relative energy $\Delta$\textit{E} with respect to dimer T (K)}} \\
    \cline{2-5}
    \textbf{} & \textbf{T $\rightarrow$ Z} & \textbf{T $\rightarrow$ X} & \textbf{T $\rightarrow$ H} & \textbf{T $\rightarrow$ L} \\
    \hline
    PZ & 9.64 & 147.77 & 166.99 & 2.45 \\
    Xa & 7.05 & 96.51 & 110.43 & 21.07 \\
    BLYP-D3 & 5.84 & 36.13 & 37.09 & 31.72 \\
    PBEsol & 4.29 & 38.91 & 46.01 & 36.05 \\
    PW91 & 3.96 & 37.46 & 45.62 & 35.77 \\
    PBE025 & 3.54 & 28.91 & 35.83 & 38.12 \\
    PBE & 3.49 & 35.46 & 42.80 & 33.02 \\
    PBE048 & 3.41 & 24.02 & 30.06 & / \\
    OEP48 & 3.21 & 21.59 & 27.22 & / \\
    RPAx[PBE] & 3.19 & 22.16 & 28.67 & / \\
    RPAx[OEP48] & 3.06 & 22.21 & 27.86 & 34.68 \\
    LRDMC & 2.97 & 26.97 & 33.72 & 38.29 \\
    X3LYP & 2.69 & / & / & / \\
    vdW-DF2 & 2.50 & 32.46 & 39.75 & 29.92 \\
    RPA[PBE] & 2.39 & 14.90 & 19.40 & / \\
    RPA[OEP42] & 2.32 & 15.41 & 19.92 & 25.90 \\
    B3LYP-D3 & 2.30 & 32.60 & 39.77 & 34.80 \\
    vdW-DF & 2.03 & 16.90 & 22.92 & 30.96 \\
    revPBE & 1.48 & 12.20 & 15.70 & 18.78 \\
    HF & 1.01 & / & / & / \\
    B3LYP & 0.13 & / & / & / \\
    BP & / & / & / & / \\
    BLYP & / & / & / & / \\
    \toprule
\end{tabular}

\end{table}

\newpage

\begin{figure*}[hbt!]
\centering
\includegraphics[scale=0.90,angle=0]{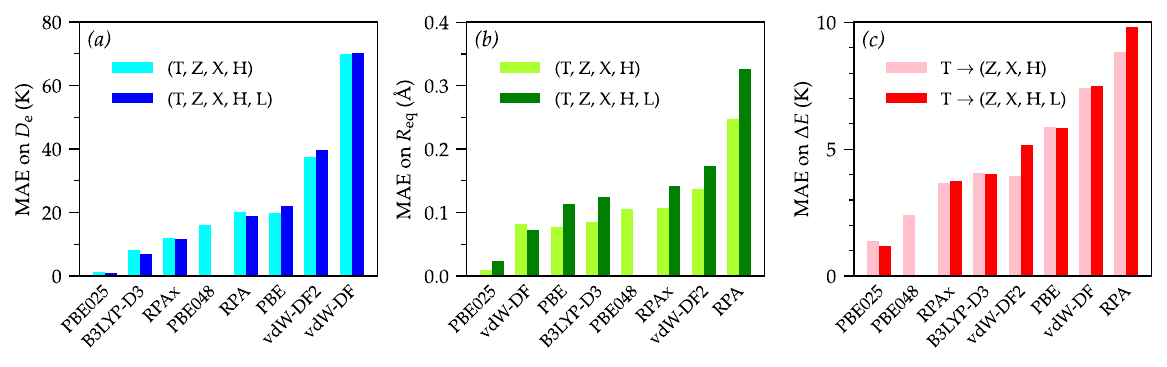}
\caption{Mean absolute error (MAE) of \textit{(a)} the dissociation energy \textit{D}$_{\rm e}$ of the dimers, \textit{(b)} the intermolecular bond distance at equilibrium geometry $R_{\rm eq}$, and \textit{(c)} the relative energy $\Delta \rm{\textit{E}}$ with respect to the T configuration. The comparison is drawn with LRDMC data. MAE values are given both with and without consideration of the L-shaped dimer. Since PBE048 does not predict the L-shaped dimer to be stable (see Tables \ref{tab:Table2_SM}, \ref{tab:Table3_SM} and \ref{tab:Table4_SM}), we have omitted the PBE048 data corresponding to this geometry.}
\label{fig:Figure5_SM}
\end{figure*}

\begin{figure}[H]
\begin{center}
\includegraphics[scale=0.80,angle=0]{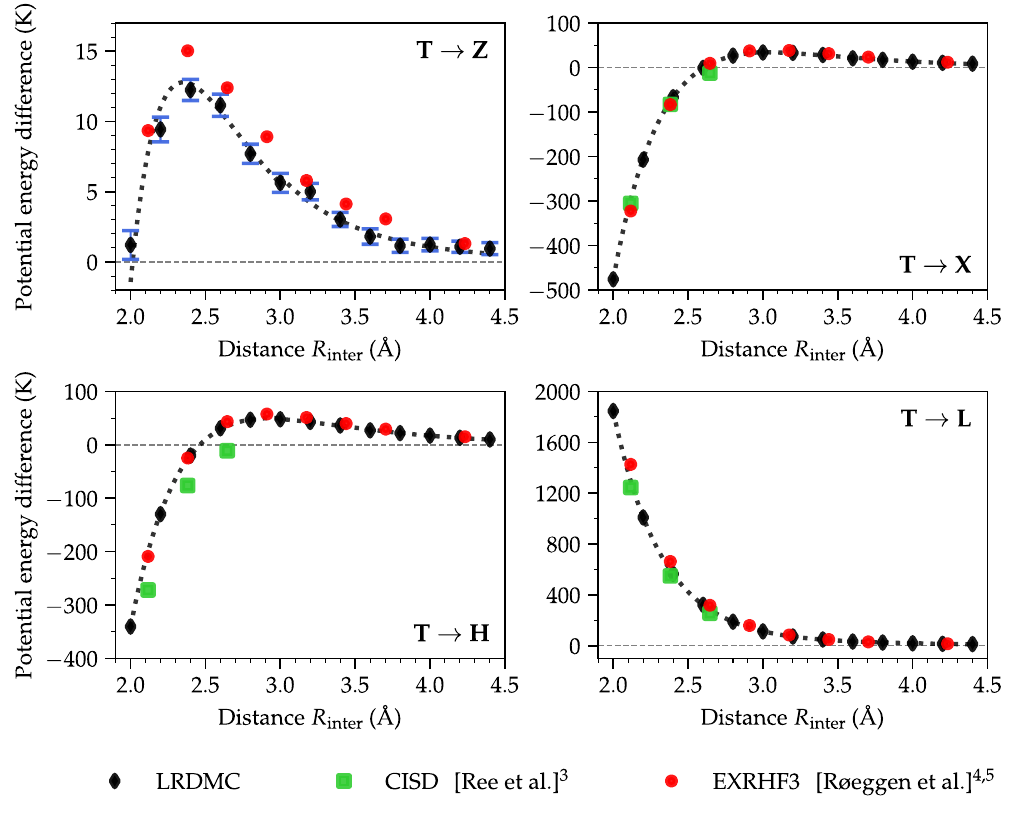}
\caption{\label{fig:Figure6_SM} Evolution of the potential energy difference between the T-shaped dimer and the other four orientations at same intermolecular distance $R_{\rm inter}$. The LRDMC raw data (black diamonds) have been fitted with Morse potentials (black dotted curves). They are compared to the CISD results obtained by Ree et al.$^{3}$ (green squares) and to the results obtained by R$\o$eggen et al. using the extended geminal EXRHF3 model with a [8s, 4p, 2d] uncontracted Gaussian basis set$^{4,5}$ (red circles). In the EXRHF3 model the double-pair correlation is approximated by a CI expansion within an orbital space comprising 36 spatial orbitals, made of 2 occupied orbitals, 8 virtual natural orbitals (NOs) per geminal for intra-pair correlation and 18 dispersive virtual NOs for inter-pair correlation. The estimated error made by this truncation is about 2K.$^{5}$ The standard deviation of LRDMC calculations remains below 1.2K for every data points. For reasons of visibility, it is only represented for data points of the T $\rightarrow$ Z panel (blue square brackets).}
\end{center}
\end{figure}

\begin{figure}[ht!]
\begin{center}
\includegraphics[scale=0.55,angle=0]{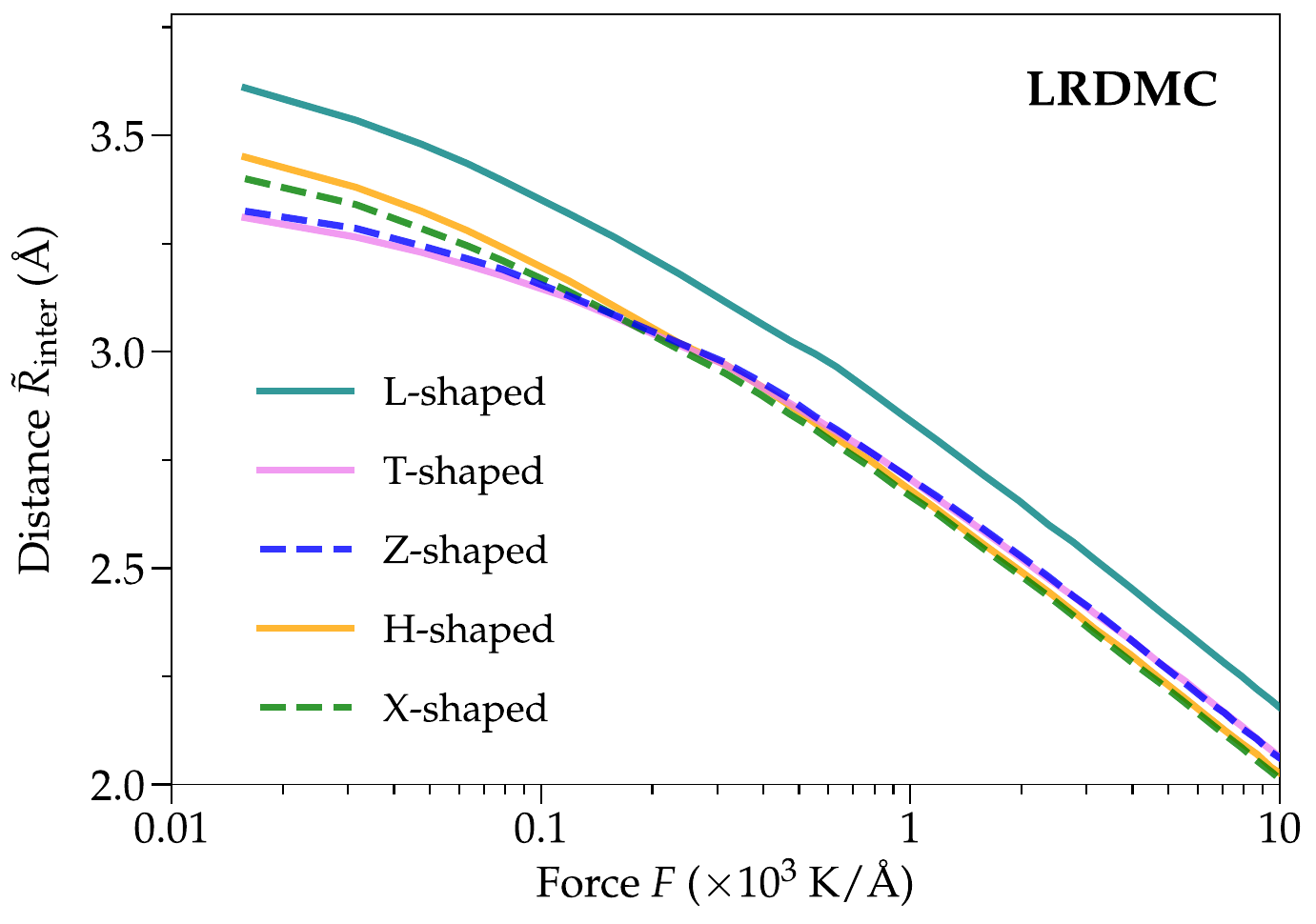}
\caption{\label{fig:Figure7_SM} Evolution of the intermolecular distance under the action of a compressive force for five different (H$_2$)$_2$ configurations according to LRDMC.}
\end{center}
\end{figure}

\vspace{310pt}

\begin{figure}[H]
\begin{center}
\includegraphics[scale=0.55,angle=0]{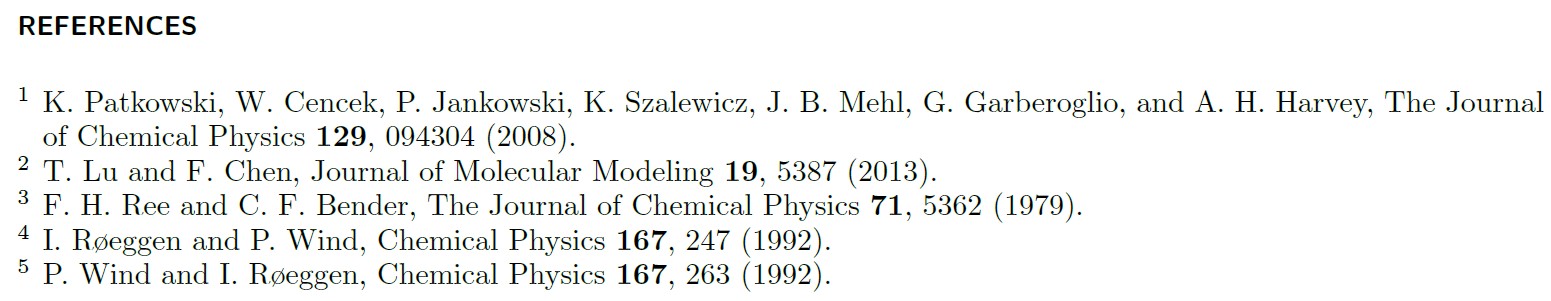}
\end{center}
\end{figure}

\end{document}